\documentclass{aa}
\usepackage[dvips]{graphicx}
\def\kms{km~s$^{-1}$}
\def\Vlsr{$V_{\rm LSR}$}
\def\HII{H{\sc ii}}
\def\wat{H$_2$O}

\def\HII{H{\sc ii}}

\def\kms{\mbox{km~s$^{-1}$}}

\def\Vlsr{$V_{\rm LSR}$}

\def\nodata{$\cdot\cdot\cdot$}

\def\h13co{H$^{13}$CO$^+$}
\def\cc34s{CC$^{34}$S}
\def\n2h{N$_2$H$^+$}

\def\13co{$^{13}$CO}
\def\c18o{C$^{18}$O}

\def\pbeam{beam$^{-1}$}

\def\i20{IRAS 20050+2720}
\def\lesssim{\mathrel{\hbox{\rlap{\hbox{\lower4pt\hbox{$\sim$}}}\hbox{$<$}}}}
\def\gtrsim{\mathrel{\hbox{\rlap{\hbox{\lower4pt\hbox{$\sim$}}}\hbox{$>$}}}}

\begin{document}
   \title{Proper Motion of H$_2$O Masers in IRAS 20050+2720 MMS1:~An AU Scale Jet Associated with An Intermediate-Mass Class 0 Source}

   \author{Ray S. FURUYA\inst{1},
           Yoshimi KITAMURA\inst{2},\\
           Alwyn WOOTTEN\inst{3}, Mark J. CLAUSSEN\inst{4} \and Ryohei KAWABE\inst{5}\\
          }

   \offprints{R. S. Furuya}

   \institute{Division of Physics, Mathematics, and Astronomy, California Institute of Technology, 
             MS 105-24, 1201 East California Boulevard, Pasadena, CA 91125, U.S.A.\\
              \email{rsf@astro.caltech.edu},
         \and
             Institute of Space and Astronautical Science,
             Japan Aerospace Exploration Agency,
              Yoshinodai 3-1-1, Sagamihara, Kanagawa 229-8510, Japan\\
              \email{kitamura@pub.isas.jaxa.jp}
         \and
              National Radio Astronomy Observatory,
              520 Edgemont Road, Charlottesville, VA 22903, U.S.A.\\
              \email{awootten@nrao.edu}
         \and
              National Radio Astronomy Observatory,
              1003 Lopezville Road, Socorro, NM 87801, U.S.A.\\
              \email{mclausse@nrao.edu}
         \and
             National Astronomical Observatory,
              Osawa 2-21-1, Mitaka, Tokyo 181-8588, Japan\\
              \email{kawabe@nro.nao.ac.jp}
             }

\date{Received 2003 August 13; accepted 2005 March 26}
\titlerunning{Proper Motion of H$_2$O Masers in IRAS 20050+2720 MMS1}
\authorrunning{Furuya et al.}

   \abstract{
We conducted a 4 epoch 3 month VLBA proper motion study 
of H$_2$O masers toward an intermediate-mass class 0 source \i20\ MMS1
($d=$700 pc).
The region of \i20\ contains at least 3 bright young stellar objects 
at millimeter to
submillimeter wavelengths and shows three pairs of CO outflow lobes:
the brightest source MMS1, 
which shows an extremely high velocity (EHV) wing emission,
is believed to drive the outflow(s).
From milli-arcsecond (mas) resolution VLBA images, 
we found two groups of \wat\ maser spots
at the center of the submillimeter core of MMS1.
One group consists of more than $\sim 50$ intense maser spots;
the other group consisting of several weaker maser spots
is located at 18 AU south-west of the intense group.
Distribution of the maser spots in the intense group shows an arc-shaped structure
which includes the maser spots that showed a clear velocity gradient.
The spatial and velocity structures of the maser spots in the arc-shape 
did not significantly change through the 4 epochs.
Furthermore, we found a relative proper motion between the two groups.
Their projected separation increased by 1.13$\pm$0.11 mas over the 4 epochs 
along a line connecting them
(corresponding to a transverse velocity of 14.4 km s$^{-1}$).
The spatial and velocity structures of the intense group and the
relative proper motions strongly suggest that the maser emission is 
associated with a protostellar jet.
Comparing the observed LSR velocities with calculated radial velocities
from a simple biconical jet model, we conclude that the most of the maser 
emission are likely to be associated with an accelerating biconical jet 
which has large opening angle of about $70^{\circ}$. 
The large opening angle of the jet traced by the masers would support 
the hypothesis that poor jet collimation is
an inherent property of luminous (proto)stars.
}
   \maketitle
%
\section{Introduction}
\label{s:intro}

Water maser surveys using single-dish radio telescopes toward 
intermediate- and low-mass young stellar objects (YSOs) have been extensively performed 
since the early 1990s (e.g., Claussen et al. 1996).
From a multi-epoch survey toward low-mass YSOs
(bolometric luminosity, $L_{\rm bol}<100~L_{\sun}$), 
Furuya et al.(2001, 2003) 
found that class 0 objects are favored sites for the masers: 
the detection rates are derived to be $\sim 40\%$ for class 0, while only
$\sim 4\%$ for class I.
It is known that the isotropic maser luminosity, 
$L_{\rm H_2O}$, correlates well with 
the bolometric luminosity of the source 
(Wilking et al. 1994; Furuya et al. 2001). 
It is interesting, however, to note that the presence of the maser emission is strictly 
related to that of high-velocity outflowing gas in the case of
high-mass YSOs (Felli, Palagi \& Tofani 1992).
\cite{fur01} showed that the \wat\ maser luminosity in low-mass
stars is more closely related to the luminosity of 100 AU
scale radio jets rather than the mechanical luminosity of larger scale
CO outflows. In fact, VLA observations have revealed that
the masers tend to be distributed within several hundred AU of the central 
stars (e.g., Wootten 1989). 
Although some H$_2$O masers are reported to be associated with protostellar disks 
(e.g., Fiebig et al. 1996; Torrelles et al. 1998; Seth, Greenhill, \& Holder 2002), 
high resolution VLBI observations have demonstrated
the presence of knots and shock structures 
which are reminiscent of those of ionized jets in the larger scale
Harbig Haro objects, suggesting that in most cases the masers originate 
in shocks produced by jets from protostars 
(e.g., Claussen et al. 1998; Furuya et al. 2000).
There are a few published VLBI water maser studies of the jets and outflows from
intermediate-mass young stellar objects (Patel et al.~2000; Seth et al. 2002). 
In order to extend our knowledge of \wat \ masers in 
intermediate-mass YSOs, we have conducted multi-epoch VLBA observations
of \wat \ masers towards the intermediate-mass YSO IRAS 20050$+$2720~
($d=$700 pc).\par

\section{IRAS 20050$+$2720}
\label{s:i20intro}

IRAS 20050$+$2720 is surrounded by a large cluster of 
low-mass stars
(Chen et al. 1997; Wilking et al. 1989) 
and has a luminosity in the IRAS bands of $388~L_{\odot}$
(Molinari et al. 1996).
The IRAS source has been categorized as a luminous 
class 0 protostar in the early compilation (\cite{bac96}).
Recent SCUBA imaging (Chini et al. 2001) revealed the presence of a bright central object 
(IRAS 20050$+$2720 MMS1) and two associated objects located $\sim 2'$ south-east.
The brightest source MMS1 is identified as the IRAS source.
Although \i20\ was not categorized as class 0 in the updated
compilation (Andr$\acute{\rm e}$, Ward-Thompson \& Barsony 2000), 
\cite{chi01} reported that the source MMS1 is very likely to be at the class 0 stage. 
This is because MMS1 shows a ratio of FIR luminosity ($L_{\rm FIR}$) and
submillimeter luminosity ($L_{\rm smm}$) 
for $\lambda\geq 350~\mu$m of
$\sim 164$ 
which satisfies one of the definitions of class 0
(Andr$\acute{\rm e}$, Ward-Thompson \& Barsony 1993;
$L_{\rm bol}/L_{\rm smm}\leq 200$).\par

Bachiller, Fuente \& Tafalla (1995; hereafter BFT95)
found three pairs of outflow lobes emanating from 
the vicinity of the source MMS1 from the
IRAM 30-m telescope CO $J=$2--1 observations.
One of the lobe pairs is a highly collimated jet with
extremely high velocity (EHV) emission whose terminal velocity 
exceeds $\sim 60$ km s$^{-1}$ with respect to the ambient cloud velocity.
The presence of the EHV outflow suggests that the driving source of the EHV outflow
is in its most powerful outflow phase.
BFT95 suggested that two or three independent outflows are emanated from 
different YSOs embedded in the cloud core,
although the driving sources have not been identified.\par

H$_2$O maser emission in IRAS 20050+2720 region was first detected by 
Palla et al. (1991),
and was subsequently observed by the Arcetri group 
(Brand et al. 1994; 
Palumbo et al. 1994):
all of the detected emission was seen around the cloud velocity 
($V_{\rm LSR}=6$ km s$^{-1}$).
Using the Nobeyama 45-m telescope, Furuya et al. (2003) 
newly detected EHV maser emission at 
$V_{\rm LSR}=-91$ km s$^{-1}$.
The EHV emission was blueshifted with respect to the cloud velocity,
while no high velocity emission was detected on the redshifted side.
This source also showed weak, blueshifted, intermediate high velocity 
(IHV) components
at $V_{\rm LSR}=-24$ and $-36~\rm km~s^{-1}$ in 1998 February.
In 1999, we carried out VLA observations and
found that all of the maser emission was located within $\sim 5''$ 
($\sim$350 AU) from the source MMS1.
Our VLA observations revealed that 
the EHV emission is located exactly at the JCMT position for source MMS1, 
while the low-velocity components around the cloud velocity are located to
$4.50\arcsec$ west and $0.60\arcsec$ north from the 
EHV emission (Furuya et al. 2003).  
Suspecting multiplicity of CO outflows and the fact that the SCUBA beam 
($8.3''$ at 450 $\mu$m; Chini et al. 2001) is larger 
than the separation of the two maser components, 
we made high resolution continuum images of the region with the OVRO mm-array.
In order to investigate the detailed structure of the masers,
we carried out extremely high angular resolution VLBA observations.
\par

\begin{table*}[ht*]
\begin{center}
\newcommand{\lw}[1]{\smash{\lower2.ex\hbox{#1}}}
\caption{Summary of VLBA H$_2$O Maser Observations}
\label{tbl:vlbaobs}
\begin{tabular}{lcccccccccc}
\hline\hline
       & & \multicolumn{4}{c}{MAIN Field: Low-Velocity Emission} & & \multicolumn{4}{c}{Sub-Field$^b$ : EHV Emission} \\
\cline{3-6}\cline{8-11}   
Epoch$^a$  & & Number of & $\theta_{\rm maj}\times\theta_{\min}$ & P.A. & Sensitivity$^c$ & & 
           Number of & $\theta_{\rm maj}\times\theta_{\min}$ & P.A. & Sensitivity$^c$ \\
           & & Antennas  & (mas$\times$mas) & (deg) & (mJy beam$^{-1}$) & & 
               Antennas  & (mas$\times$mas) & (deg) & (mJy beam$^{-1}$) \\ \hline
I   & & 10$^d$ & 0.78$\times$0.40 & $-2$ & 3.4 & & 5$^f$ & 2.8$\times$1.4 & $+$29 & 6.2 \\
II  & & 9$^e$  & 1.1$\times$0.51 & $+26$ & 3.7 & & 5$^f$ & 2.8$\times$1.5 & $+$29 & 4.7 \\
III & & 10$^d$ & 0.95$\times$0.40 & $-14$ & 2.8 & & 5$^f$ & 2.9$\times$1.6 & $+$24 & 4.8 \\
IV  & & 10$^d$ & 1.1$\times$0.39 & $-14$ & 3.5 & & 5$^f$ & 2.9$\times$1.6 & $+$27 & 6.9 \\
\hline
\end{tabular}
\end{center}
$(a)$ The 4 epochs are 1999 April 1, May 5, June 5 and July 4.
$(b)$ The following data are common for the Sub-Fields 1 and 2 (see $\S$\ref{sss:ehvsearch}).
$(c)$ An RMS image noise level with a velocity resolution of 0.2 km s$^{-1}$.
$(d)$ All 10 VLBA stations.
$(e)$ Except the North Liberty station.
$(f)$ Fort Davis, Los Alamos, Pie Town, Kitt Peak and Owens Valley.
\end{table*}

   \begin{figure*}[ht*]
   \centering
   \includegraphics[width=6.7cm,angle=-90,clip]{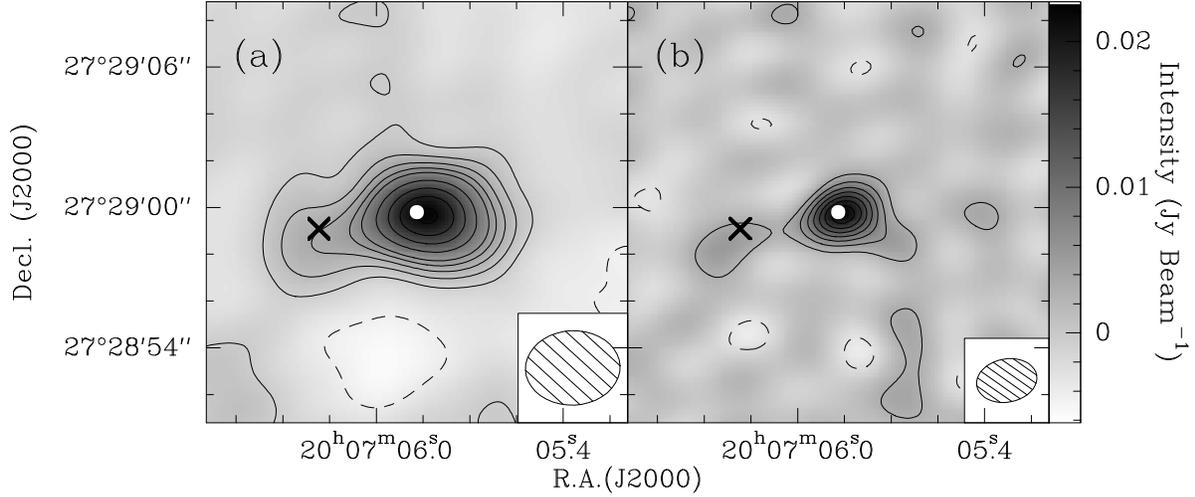}
      \caption{
Continuum emission maps taken with the OVRO millimeter array
toward IRAS 20050+2720 MMS1 with
(a) natural weighting ($left$) and (b) uniform weighting ($right$).
The solid contours start at $+2\sigma$ level with $+2\sigma$ interval and 
the dashed contours are $-2\sigma$ intervals started at $-2\sigma$ level.
The image noise levels are
0.78 and 1.5 mJy beam$^{-1}$ for the panels (a) and (b), respectively.
The white dot and the eastern cross indicate the positions of the low-velocity and EHV
H$_2$O maser emission (see $\S$\ref{sss:ehvsearch}).
The sizes of the synthesized beams are presented in the bottom right corners.
              }
         \label{fig:3mmcont}
   \end{figure*}

\section{Observations}
\label{s:obs}

\subsection{3 mm Continuum Emission Observations with OVRO mm-Array}

Aperture synthesis observations of continuum emission at 3 mm were 
carried out using the six-element Owens Valley Radio Observatory (OVRO)
Millimeter Array from 2002 December to 2003 February with the H and E array configurations. 
The phase tracking center was 
$\alpha$(J2000)=20$^{\rm h}$07$^{\rm m}$5\farcs87,
$\delta$(J2000)=27\degr28\arcmin59\farcs 80.
The field of view (FOV) was 65\arcsec.
All of the element antennas are equipped with SIS receivers having system noise
temperatures in double-sideband of 200 K toward the zenith at 93~GHz.
We tuned the 3 mm SIS receiver at the frequencies of N$_2$H$^+$ (1--0) line
(93.173 GHz) for upper sideband and
H$^{13}$CO$^+$ (1--0) line (86.754 GHz ) for lower side band.
A detailed presentation of the results of the molecular line emission will be 
published elsewhere (R. S. Furuya et al. 2005, in preparation).
The Continuum Correlator was configured for both the
sidebands, with a total bandwidth of 3 GHz.
We used 3C\,454.3 and 3C\,84 as a passband calibrator and J2025+337
as a phase and gain calibrator.
Flux density of J2025$+$337 was measured by comparison of that of Uranus:~
it was stable in the range
from 1.2 to 1.5~Jy during the observation period. 
The overall flux uncertainty is about 20\%. 
The data calibration was done using the originally developed software at OVRO,
 and the image construction was performed using the AIPS package of the NRAO. 
After merging the data in both the sidebands, 
we constructed continuum emission images
with two beam weightings.
Synthesized beam sizes were $4.''04\times 3.''18$ with natural weighting and
$2.''60\times 1.''85$ with uniform weighting.
The 1$\sigma$~rms noise levels for the continuum emission maps were 
0.78~mJy~beam$^{-1}$ for the former and 1.5~mJy~beam$^{-1}$ for the latter.\par

\subsection{H$_2$O Maser Observations with VLBA}

VLBI observations of the H$_2$O maser emission in IRAS 20050+2720 MMS1
were carried out using all 10 antennas of the Very Long Baseline Array (VLBA) of the 
NRAO\footnote{The National Radio Astronomy Observatory (NRAO) is operated 
by Associated Universities, Inc., under cooperative agreement with the 
National Science Foundation}
on 1999 April 1, May 5, June 5, and July 4
(hereafter epochs I, II, III and IV, respectively).
For epoch II, however, we could not use the antenna at North Liberty.
All of the data were obtained for 8-hour integration in each epoch.
We used a frequency setup of the 8 MHz IF bandwidth mode with 512 channels,
which provides a velocity resolution of 0.2 km s$^{-1}$ at 22.235077 GHz.
This frequency setup covers the range of $V_{\rm LSR}=-96.1$ to 
$+10.5$ km s$^{-1}$: this velocity coverage is sufficient to detect all of the 
maser emission previously detected.\par

The data were correlated at the NRAO Array Operation Center (Socorro, New Mexico).
We adopted a correlator averaging time of 2.16 sec to
obtain $\sim 0.8''$ radius FOV for the baseline of $\sim 4000$ km. 
Data calibration and image construction were performed using the AIPS package developed by the NRAO.
We used two bright quasars 3C345 and 3C454.3 to determine delay 
and fringe rates as well as to calibrate bandpass response.
In the next section, we present the image construction, 
identification of the maser emission
and further analysis together with the results.\par

\section{Results and Analyses}

\subsection{Relation between Millimeter Continuum Emission and Masers}
\label{ss:mmcont}

Figure \ref{fig:3mmcont} presents the OVRO continuum emission
maps together with the positions of the H$_2$O masers.
There is a distinct continuum emission peak toward 
the low-velocity and EHV maser emission.
To plot the absolute position of the low-velocity masers, 
we adopted the results from the VLA measurements by Furuya et al. (2003) 
and used the position offset of the EHV emission
obtained in $\S$\ref{sss:ehvsearch}.
Clearly the millimeter continuum emission is associated with
the low-velocity maser emission.
The peak of the continuum
emission is slightly shifted with respect to the maser emission.
We believe that this positional shift is real considering the baseline accuracies of
VLA and OVRO array 
and angular separations between the calibrators and the source. 
It is noteworthy that the continuum emission in the
natural weighting map (Figure \ref{fig:3mmcont}a) 
is elongated to the east, namely, toward the EHV emission.
In the uniform weighting map (Figure \ref{fig:3mmcont}b), 
one may notice that there is a
weak emission peak close to the EHV maser emission:
this fact might suggest that there are at least two YSOs in this region.\par

We estimate molecular hydrogen mass ($M_{\rm H_2}$) of the
core from the total flux density ($F_{\nu}$) of the continuum emission
assuming that the emission is thermal radiation from dust grains.
We used a relation 
$M_{\rm H_2}=F_{\nu}D^2/(\kappa_{\nu}B_{\nu}(T_{\rm d}))$
where
$\kappa_{\nu}$ is the mass opacity coefficient of the dust, 
$T_{\rm d}$ is the dust temperature and
$B_{\nu}(T_{\rm d})$ is the Planck function.
The value of $\kappa_{\nu}$ at 3 mm is calculated with the
usual form of $\kappa_{\nu}\propto \kappa_0\nu^\beta$.
In order to keep consistency with the previous JCMT measurements by Chini et al. (2001),
we used the same $\kappa_0$ of 0.003 cm$^2$ g$^{-1}$ at 231 GHz,
$\beta$ of 1.4, and $T_d$ of 34 K.
We obtained $M_{\rm H_2}$ of 0.007$M_{\sun}$ for $F_{\nu}=25.6$ mJy
by integrating the emission inside the 3$\sigma$ contour.
Note that the derived mass is a lower limit because interferometric 
observations do not receive the whole flux from a source 
due to the lack of short spatial frequency data.
In fact, the resultant projected baseline length of our OVRO observations
ranged from 6.2 to 75 k$\lambda$, which will 
miss 50\% of the flux from structures extending more than 
$16''$ (0.05 pc at $d=$700 pc)(see Wilner \& Welch 1994).

\subsection{Maser Emission Search in the VLBA Data:~Low and Extremely High Velocity Emission}

In the following, we present results and analyses from the VLBA
observations of the \wat \ masers.\par

First, we carried out fringe-frequency analysis 
(e.g., Walker 1981; Walker, Matsakis \& Garcia-Barreto 1982)
to cover a large FOV of $\sim 3''$ which is 3 orders of magnitude 
larger than the VLBA fringe-spacings.
The purpose of the fringe-frequency analysis was to search for 
maser spots which were not excited during the VLA observations in 1999 February
(Furuya et al. 2003).
As expected from the VLA observations, we confirmed that 
the distribution of the low-velocity masers is sufficiently compact to 
perform standard Fourier synthesis.
The EHV emission was too weak to be detected with the fringe-frequency analysis.\par

Subsequently, we performed self-calibration using a
strong ($\sim 45$ Jy measured by Furuya et al.~2003) and 
point-like maser spot identified at 
$V_{\rm LSR}=+1.6$ km s$^{-1}$ as a model.
In the self-calibration procedure, we solved the time variation of the 
complex gain for phase weighted by amplitude.

\subsubsection{Main Field:~Low-Velocity Emission}

Applying the solutions from the self-calibration procedure, 
we carried out image construction toward the low-velocity 
emission (hereafter we refer to it as the MAIN Field): 
the image area has $\sim 0.23''$ size divided into 9 fields each of which
has 77 milli-arcsecond (mas) size and 
512$\times$512 pixels with a cell size of 0.15 mas.
We believe that the area size of $\sim 0.23''$ was sufficient to search for
maser emission considering the absolute position accuracy of 
the VLA observations ($\S$\ref{s:i20intro}).
Subsequent to this coarse search, 
we constructed a final image of 41 mas size for the area 
where the low-velocity emission was detected,
with a smaller cell size of 0.08 mas.
The image noise level per velocity channel was typically 3 mJy beam$^{-1}$ 
and the synthesized beam size was typically 1.0$\times$0.5 mas
(Table \ref{tbl:vlbaobs}).\par

\begin{figure}[ht]
   \centering
    \includegraphics[width=8.2cm,angle=-90,clip]{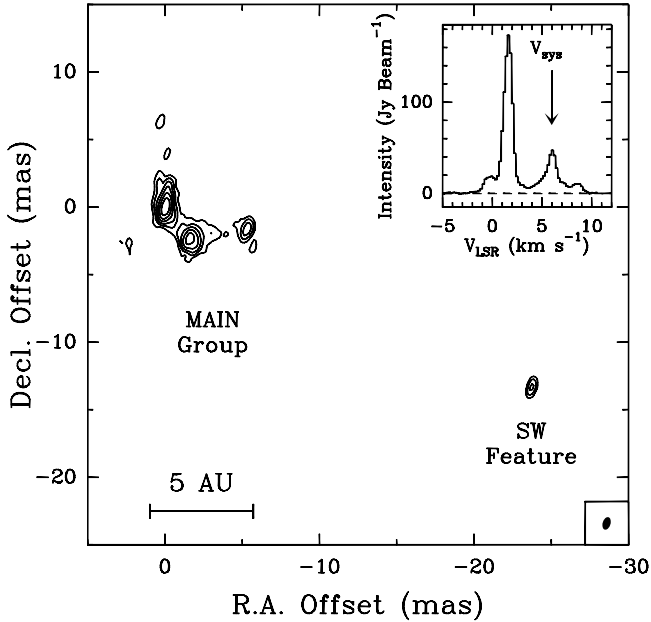}
      \caption{
Total integrated intensity map of the H$_2$O masers in the MAIN Field of
IRAS 20050+2720 MMS1 obtained at the 3rd epoch (1999 June 5) of the VLBA observations. 
Contour levels are 10, 20, 30, 50, 100, 200, 400, 800, 1600, 3200, 6400,
12800, and 25600 times the image noise level of
12.3 mJy \pbeam \ \kms .
The inserted panel in the top right corner shows the H$_2$O maser spectrum 
obtained by integrating the emission over the presented region.
The vertical arrow shows the ambient cloud velocity ($V_{\rm sys}$) of 
$V_{\rm LSR}=6.0 $ km s$^{-1}$ estimated from 
CS $J=3$--2 line observations (BFT95).
The size of the synthesized beam is presented in the bottom right corner.
              }
         \label{fig:overall}
   \end{figure}

In Figure \ref{fig:overall}, we show a total integrated intensity map of the MAIN Field,
namely low-velocity H$_2$O masers, obtained in the Epoch III,
in which we attained the highest sensitivity among the 4 epochs.
The inserted panel presents the spectrum of the maser emission detected
in the region.
The overall distribution of the masers was similar for all 4 epochs:
there can be seen bright maser emission peaks 
at the field center (hereafter MAIN group) and an isolated emission peak 
at $\sim 25$ mas (corresponding to $\sim 18$ AU)
south-west of the MAIN group. 
Hereafter we call the latter emission as ``SW feature'' instead of 
``SW group'' because the emission showed a point-like structure.
The definition of ``feature'' will be given in $\S$\ref{ss:identification}.

\subsubsection{Extremely High Velocity Emission}
\label{sss:ehvsearch}

In addition, we searched for the EHV emission on the basis of the snapshot
VLA D-array observations in 1999 February (Furuya et al.~2003)
and 2003 January (M. Claussen, private communication):
the former observations (hereafter Sub-Field 1) showed that the
$V_{\rm LSR}=-93$ km s$^{-1}$ emission is shifted by
$+2.33\arcsec$ in R.A. and
$-0.60\arcsec$ in Decl. with respect to the low-velocity emission and
the latter observations (hereafter Sub-Field 2) showed that 
the $V_{\rm LSR}=-72$ and $-78$ km s$^{-1}$ emission is shifted by
$+4.4\arcsec$ in R.A. and
$-0.6\arcsec$ in Decl..
Since these positions are outside the FOV correlated toward the
1.6 \kms \ emission, 
we used UV data only from
south-western 5 antennas (Fort Davis, Los Alamos, Pie Town, Kitt Peak and Owens Valley)
which provide the maximum spatial frequency of $\sim 1.1\times 10^8~\lambda$.
The synthesized beam sizes were typically 2.8$\times$1.5 mas and
the image noise levels per velocity channel were typically 5 mJy beam$^{-1}$ 
(Table \ref{tbl:vlbaobs}).
Applying the solution from the self-calibration procedure above,
we constructed images toward both the Sub-Fields.

We detected the maser emission peaked at $V_{\rm LSR}=-93$ km s$^{-1}$ 
in the Sub-Field 2 where the $V_{\rm LSR}=-72$ and $-78$ km s$^{-1}$ emission 
was detected with the VLA in 2003 January.
However, we did not see any maser emission around 
$V_{\rm LSR}=-72$ and $-78$ km s$^{-1}$ toward the two Sub-Fields.
The detected $-93$ km s$^{-1}$ emission shows a point-like structure
located at $\sim$4184 mas east and $\sim$709 mas 
south with respect to the 
$V_{\rm LSR}=1.6$ km s$^{-1}$ 
emission peak with which we performed the self-calibration.

   \begin{figure*}[ht*]
   \centering
      \includegraphics[width=14.8cm,angle=0,clip]{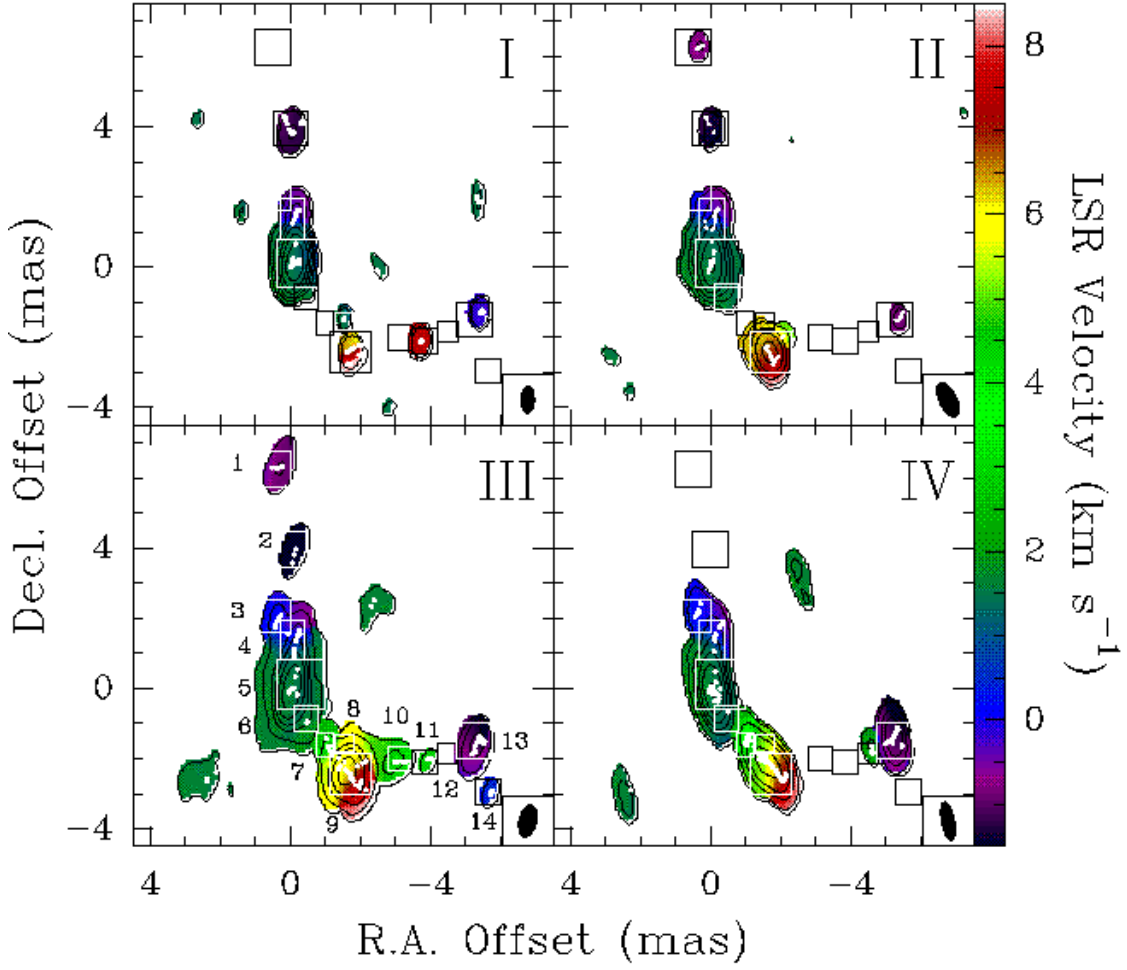}
      \caption{
Total integrated intensity maps (contours) overlaid with isovelocity maps
(images) for the MAIN Field of the H$_2$O masers in IRAS 20050+2720.
The contour levels are the same as those in Figure \ref{fig:overall}.
The velocity range shown here is from 
$V_{\rm LSR}=-1.5$ to 8.5 km s$^{-1}$.
The white dots present peak positions of the maser spots.
Boxes with labels indicate identified maser features.
              }
         \label{fig:overlay}
   \end{figure*}

   \begin{figure*}[ht*]
   \centering
    \includegraphics[width=6.2cm,angle=0]{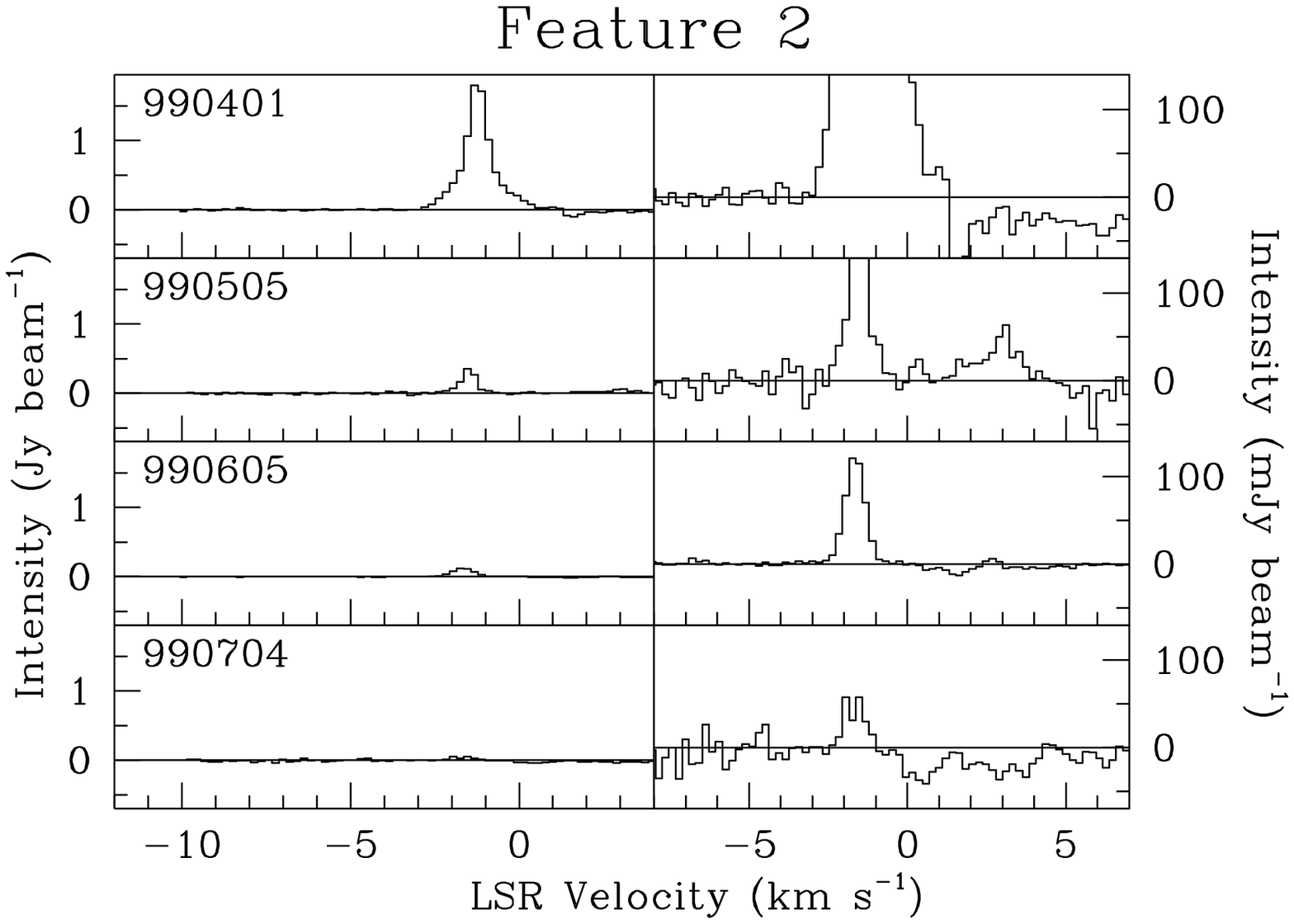}
    \includegraphics[width=6.2cm,angle=0]{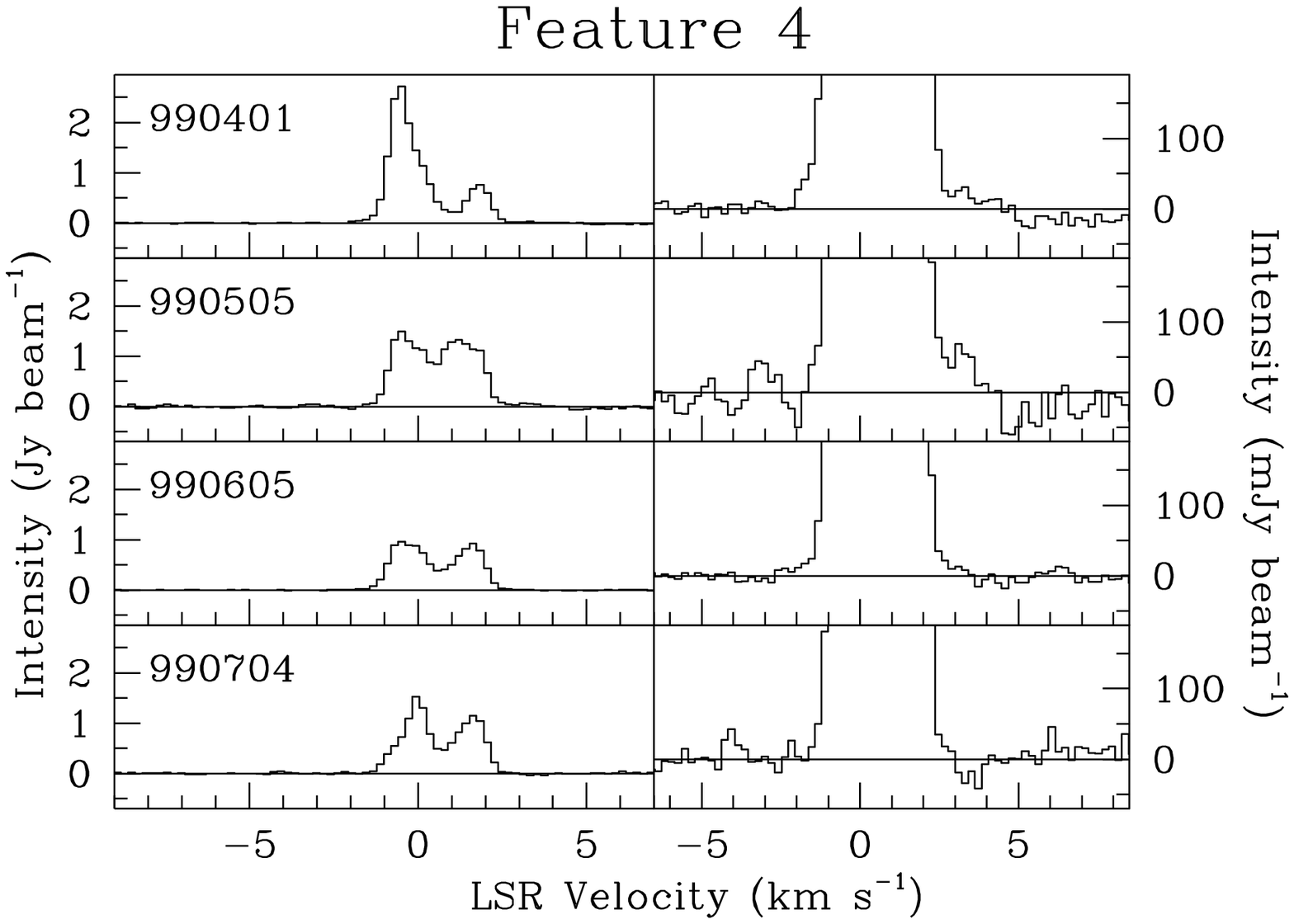} \\
    \includegraphics[width=6.2cm,angle=0]{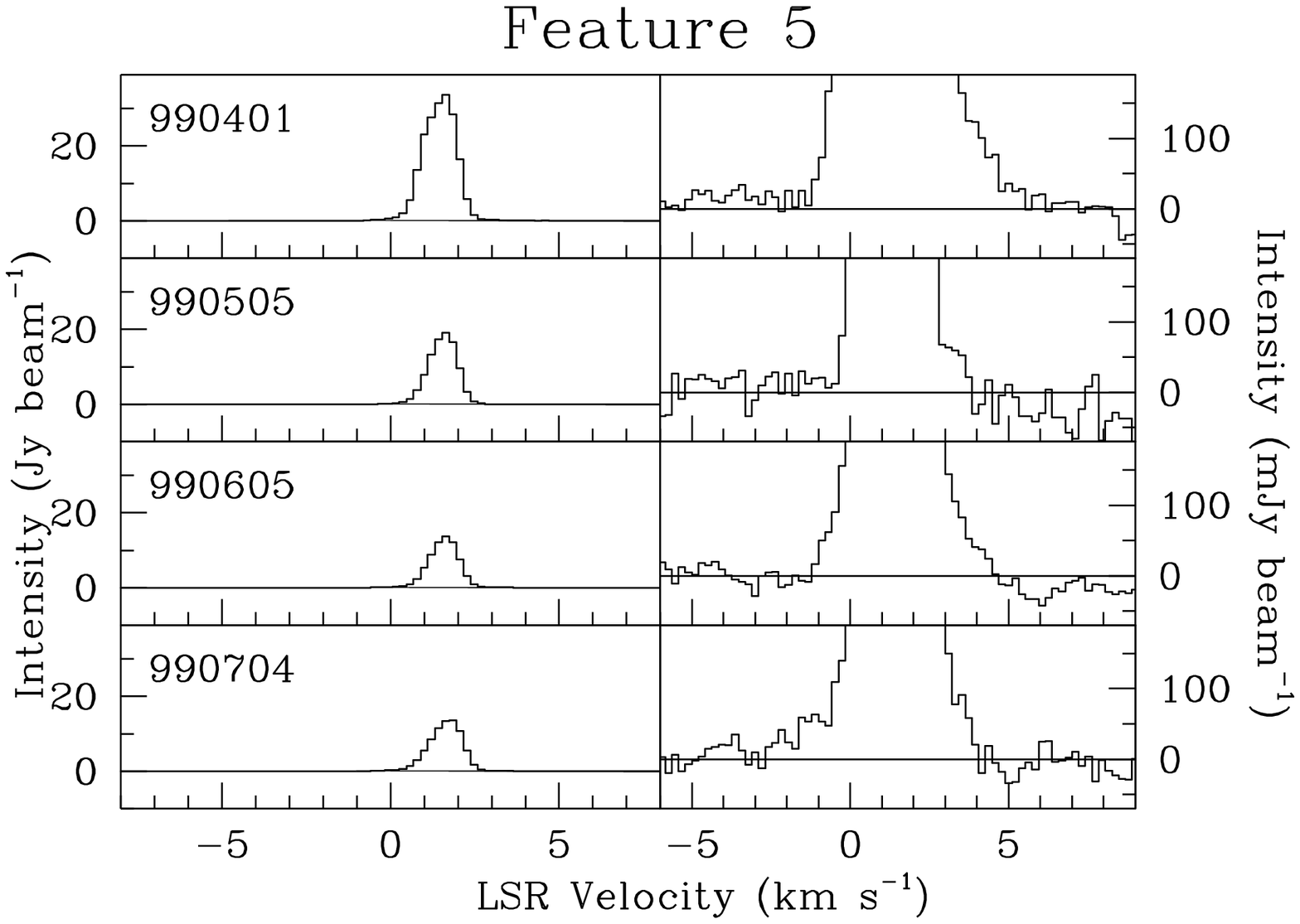}
    \includegraphics[width=6.2cm,angle=0]{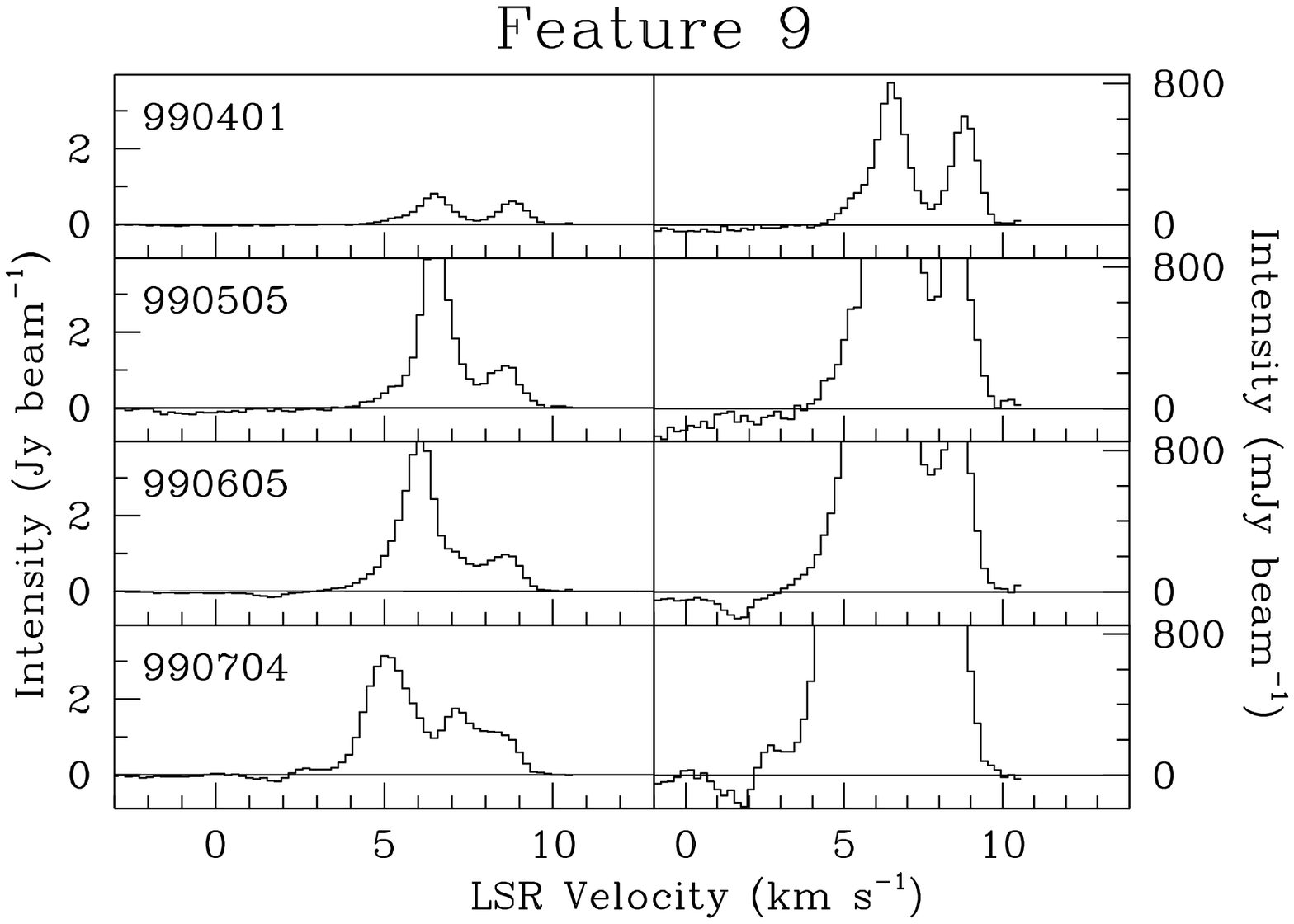} \\
    \includegraphics[width=6.2cm,angle=0]{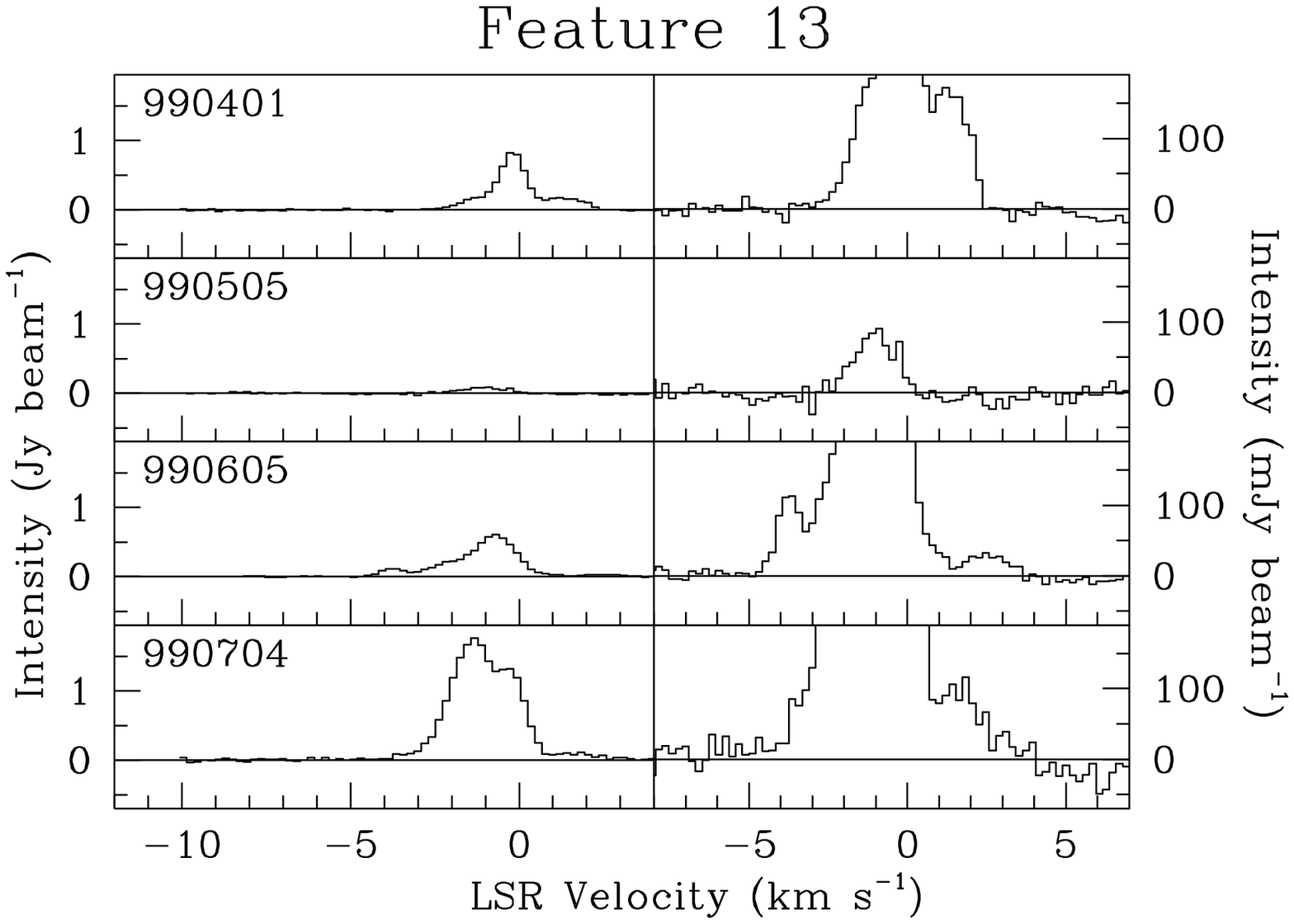}
    \includegraphics[width=6.2cm,angle=0]{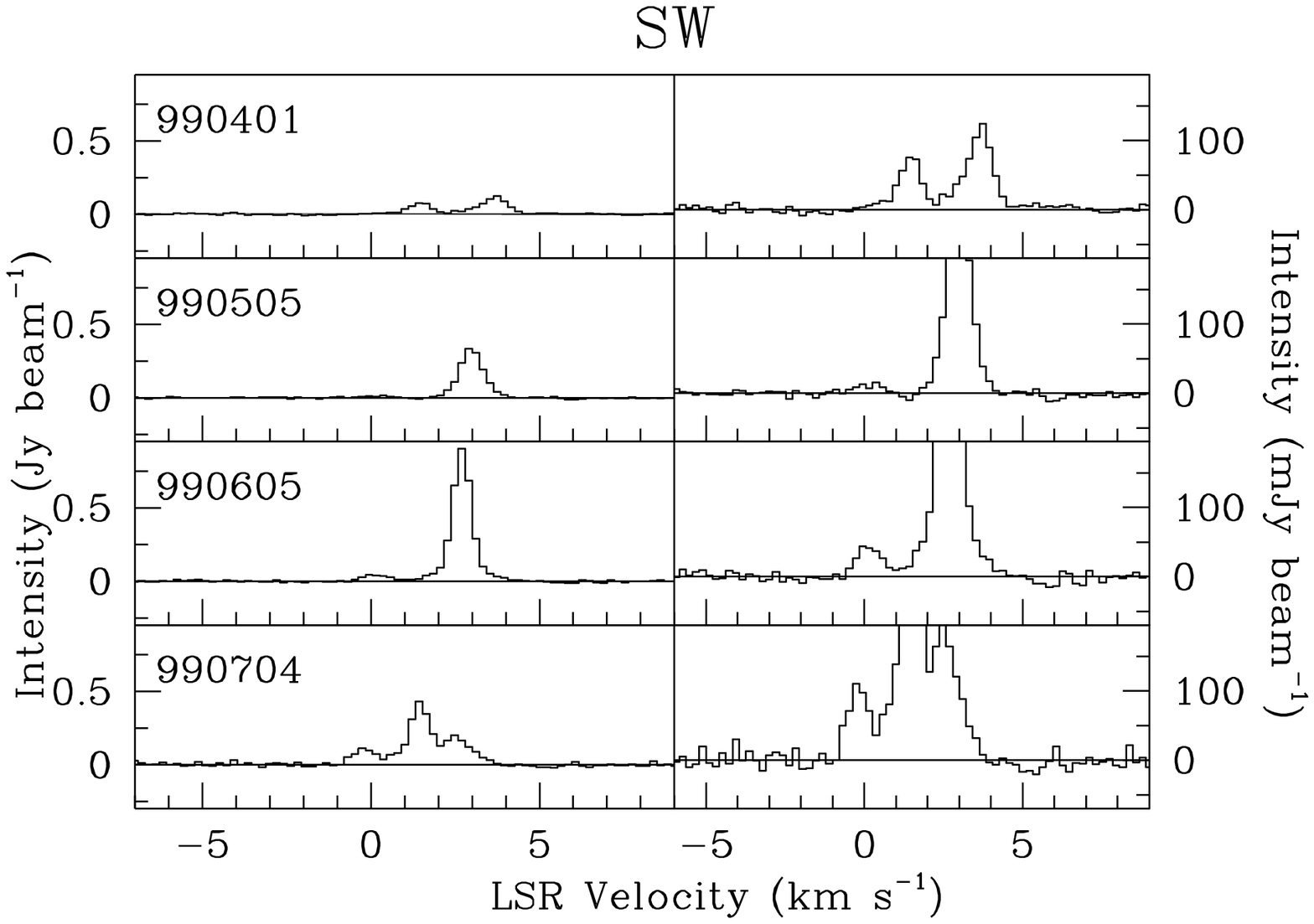} \\
    \includegraphics[width=6.2cm,angle=0]{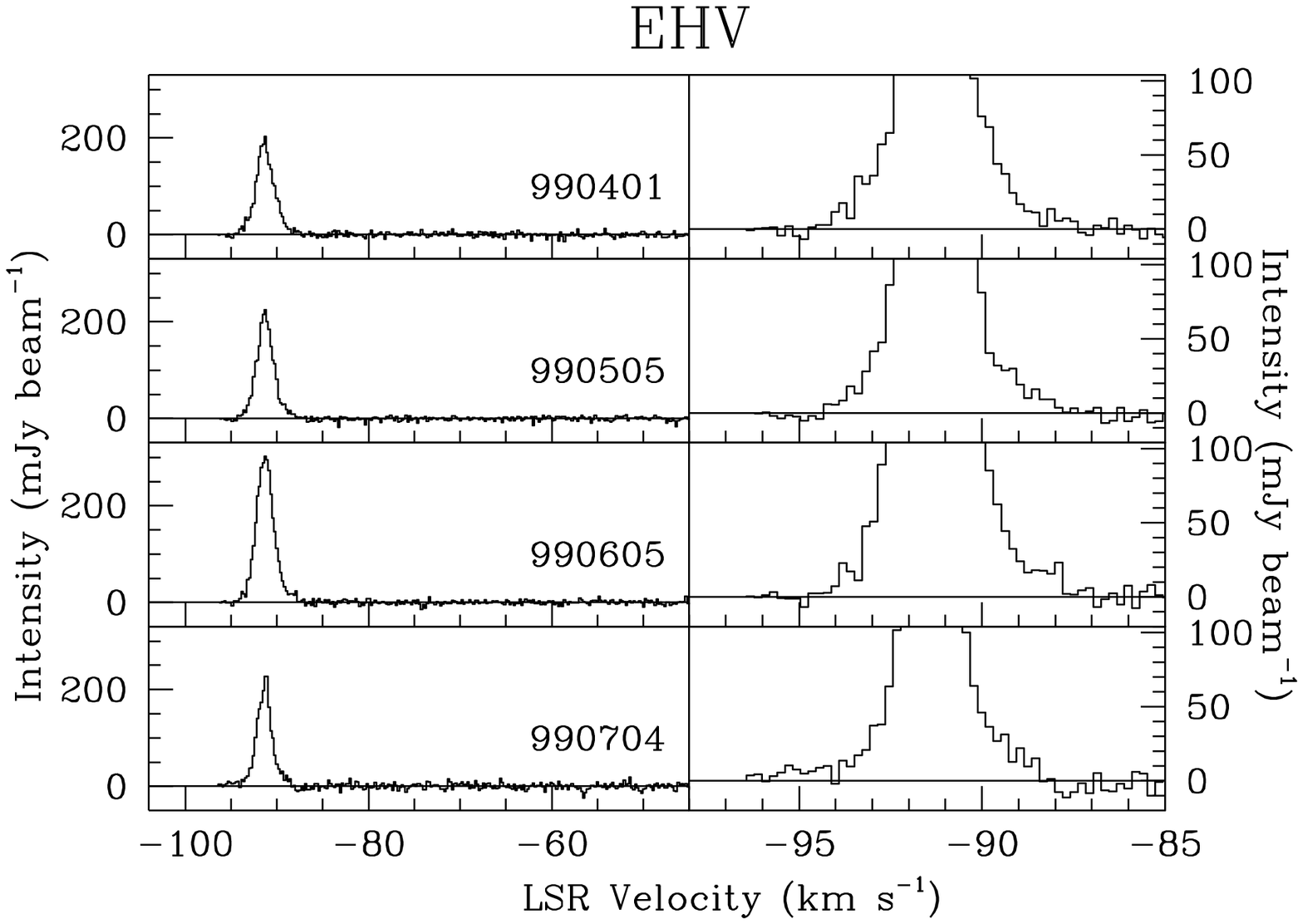}
      \caption{
H$_2$O maser spectra toward Features 2, 4, 5, 9, 13, SW and EHV
in IRAS 20050+2720 MMS1.
The ambient cloud velocity is 
$V_{\rm LSR}=6.0 $ km s$^{-1}$ (BFT95).
In each feature, the spectra in the left-hand panels are magnified in the right-hand panels.
              }
         \label{fig:sp}
   \end{figure*}

   \begin{figure*}[ht*]
   \centering 
   \includegraphics[width=14.8cm,angle=0,clip]{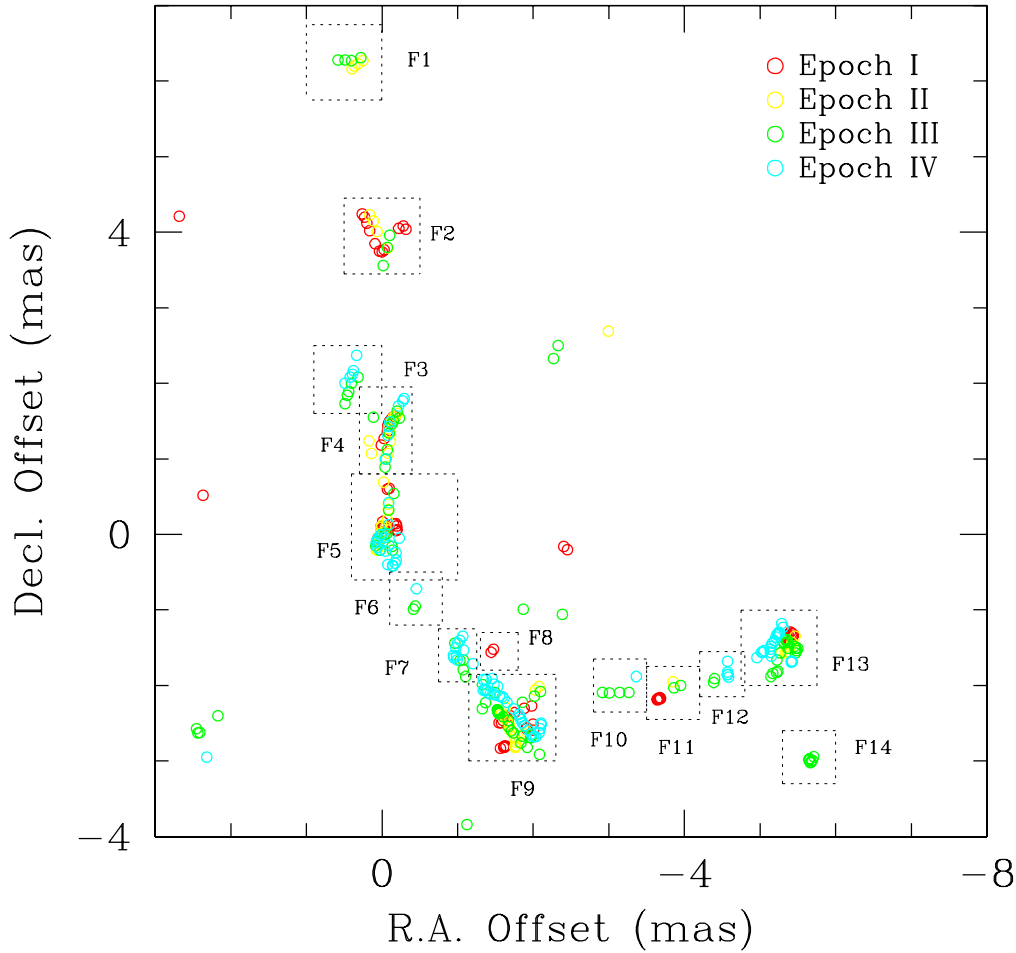}
      \caption{
Overlay of the peak positions of the low-velocity H$_2$O maser spots in the
MAIN-Field of IRAS 20050+2720 MMS1.
The colors present the observing epochs and the dashed-boxes indicate 
the maser features identified in $\S\ref{ss:identification}$.
The dashed-boxes are the same as the solid boxes in Figure \ref{fig:overlay}.
}
         \label{fig:overlay2}
   \end{figure*}


\begin{table*}[ht*]
\begin{center}
\newcommand{\lw}[1]{\smash{\lower2.ex\hbox{#1}}}
\caption{Relative Positions of the Identified H$_2$O Maser Features}
\label{tbl:posfeature}
\begin{tabular}{lrrrrrrrrrrrrrrrr}
\hline\hline
       & & \multicolumn{3}{c}{Epoch I} & & \multicolumn{3}{c}{Epoch II} & & \multicolumn{3}{c}{Epoch III} & & \multicolumn{3}{c}{Epoch IV} \\
\cline{3-5}\cline{7-9}\cline{11-13}\cline{15-17}   
Features & & $\Delta\alpha^a$ & $\Delta\delta^b$ & \multicolumn{1}{c}{$\sigma^c$} & & $\Delta\alpha$ & $\Delta\delta$ & \multicolumn{1}{c}{$\sigma$} & & $\Delta\alpha$ & $\Delta\delta$ & \multicolumn{1}{c}{$\sigma$} & & $\Delta\alpha$ & $\Delta\delta$ & \multicolumn{1}{c}{$\sigma$} \\
       & & (mas) & (mas) & (mas) & & (mas) & (mas) & (mas) & & (mas) & (mas) & (mas) & & (mas) & (mas) & (mas) \\
\hline
2     & & $-$0.046 &    3.928 & 0.011 & &    0.049 &    4.094 & 0.014 & & $-$0.133 &    3.860 & 0.012 & & \nodata & \nodata & \nodata \\
4     & & $-$0.160 &    1.500 & 0.031 & & $-$0.124 &    1.434 & 0.010 & & $-$0.154 &    1.308 & 0.012 & & $-$0.172 & 1.510    & 0.008 \\
5     & & $-$0.133 &    0.135 & 0.008 & & $-$0.034 &    0.113 & 0.009 & & $-$0.053 & $-$0.021 & 0.018 & & $-$0.071 & $-$0.037 & 0.019 \\
9     & & $-$1.794 & $-$2.481 & 0.014 & & $-$1.679 & $-$2.436 & 0.012 & & $-$1.659 & $-$2.398 & 0.019 & & $-$1.684 & $-$2.204 & 0.010 \\
13    & & $-$5.439 & $-$1.280 & 0.010 & & $-$5.392 & $-$1.464 & 0.016 & & $-$5.349 & $-$1.600 & 0.023 & & $-$5.235 & $-$1.395 & 0.019 \\
SW    & & $-$23.11 & $-$12.89 & 0.052 & & $-$23.43 & $-$13.10 &  0.04 & & $-$23.62 & $-$13.50 & 0.03 & & $-$23.97 & $-$13.60 & 0.05 \\
EHV   & &   4184.8 & $-$709.9 & 0.29  & &   4184.3 & $-$709.9 &  0.35 & &   4183.9 & $-$709.7 & 0.18  & &  4183.6  & $-$709.4 & 0.17 \\
\hline
\end{tabular}
\end{center}
$(a)$ Right Ascension offset with respect to the 1.6 \kms \ spot,
$(b)$ Declination offset,
$(c)$ Position error
\end{table*}

\begin{table}[ht]
\begin{center}
\newcommand{\lw}[1]{\smash{\lower2.ex\hbox{#1}}}
\caption{Mean LSR-Velocities$^a$ ($V_{\rm mean}$) for the Identified Maser Features}
\label{tbl:spfeature}
\begin{tabular}{lcrrrr}
\hline\hline
\lw{Feature} & & \multicolumn{4}{c}{$V_{\rm mean}$ (km s$^{-1}$)} \\
\cline{3-6}
        & &       I &       II &      III & IV      \\ \hline
2   & &  $-$1.2 &  $-$1.1  &  $-$1.8  & \nodata \\
4   & & $-$0.39 &  $-$0.30 &  $-$0.23 &  $-$0.1 \\
5   & &     1.6 &     1.6  &     1.6  &     1.6 \\
9   & &     7.5 &     7.0  &     6.4  &     6.1 \\
13  & & $-$0.30 &  $-$1.0  &  $-$1.6  &  $-$1.1 \\
SW  & &     3.5 &     2.9  &     2.9  &     1.8 \\
EHV & & $-$91.2 & $-$91.3  & $-$91.2  & $-$91.3 \\ \hline
\hline
\end{tabular}
\end{center}
$(a)$ Errors are typically less than 0.03 \kms .
\end{table}

\subsection{Identification of Maser Spots and Features}
\label{ss:identification}

To identify maser ``spots'' from the image,
we fitted a two-dimensional elliptical Gaussian profile
to individual possible spots.
We adopted a detection threshold of a signal-to-noise ratio
(S/N) larger than 10.
In this way, we detected 30$\sim$70 maser spots at each epoch 
(Figure \ref{fig:overlay}).
We estimated a relative positional error for each spot ($\sigma_{\rm spot}$) to be
$\leq 0.02$ mas: a relative positional error of a point-like source convolved 
with a Gaussian shaped beam is given by the relation of
$\sigma=0.45~\theta_{\rm syn}/[{\rm S/N}]$ 
(\cite{con97}) where 
$\theta_{\rm syn}$ is the FWHM of the synthesized beam
($\theta_{\rm syn}=[\theta_{\rm maj}\times\theta_{\rm min}]^{1/2}$).\par

Subsequent to the identification of the maser ``spots'', 
we divided them into groups, i.e., 
spatially localized ``features'' with distinct peaks in line profiles:
we grouped the spots in adjacent velocity channels
that are distributed within one synthesized beam width.
Each feature is considered probably to represents a distinct clump of gas.
In Figure \ref{fig:overlay}, 
only Features 4, 5, 9 and 13 showed emission
over the 4 epochs among the 14 features identified in the MAIN-Field.
Feature 2 persisted from Epoch I to III,
but disappeared in the Epoch IV.
The other 9 features did not persist continuously more than 3 epochs.
Therefore, we do not consider these 9 features for our further analysis 
in proper motion measurements.
For each feature, we calculated 
an intensity-weighted mean position of the contributing maser spots:
its uncertainty is given by 
$\sigma_{\rm feature}=[\sum_{\rm spot} \sigma_{\rm spot}^2]^{1/2}$.
As summarized in Table \ref{tbl:posfeature}, the resultant positional errors were 
a few $\times$0.01 mas for the 6 features in the MAIN-Field and
a few $\times$0.1 mas for the EHV feature in the Sub-Field.\par

Finally, it is noteworthy that 
there is a velocity gradient of approximately 9 km s$^{-1}$ over 10 mas
from Features 1 to 9.

\subsection{Spectra of the Features and Their Time Variation}
\label{ss:spectra}

For each maser feature identified, we made a spectrum 
by integrating the intensity over the corresponding region.
Figure \ref{fig:sp} represents spectra of the features
listed in Table \ref{tbl:posfeature}.
Among the series of the 7 spectra, Features 2, 5 and EHV showed single-peaked
spectra while Features 4, 9, 13 and SW displayed multi-peaked spectra. 
Using these spectra, we evaluated intensity-weighted mean velocities 
($V_{\rm mean}$; Table \ref{tbl:spfeature}).
It is noteworthy that the former single-peaked features did not show any prominent 
trend of LSR-velocity change over the 4 epochs.
On the other hand, 
the latter multiple-peaked features displayed small drifts.
We discuss the velocity drifts when we assess proper motions
of the features in the next subsection.

   \begin{figure*}[ht*]
   \centering  
    \includegraphics[width=4.2cm,angle=0,clip]{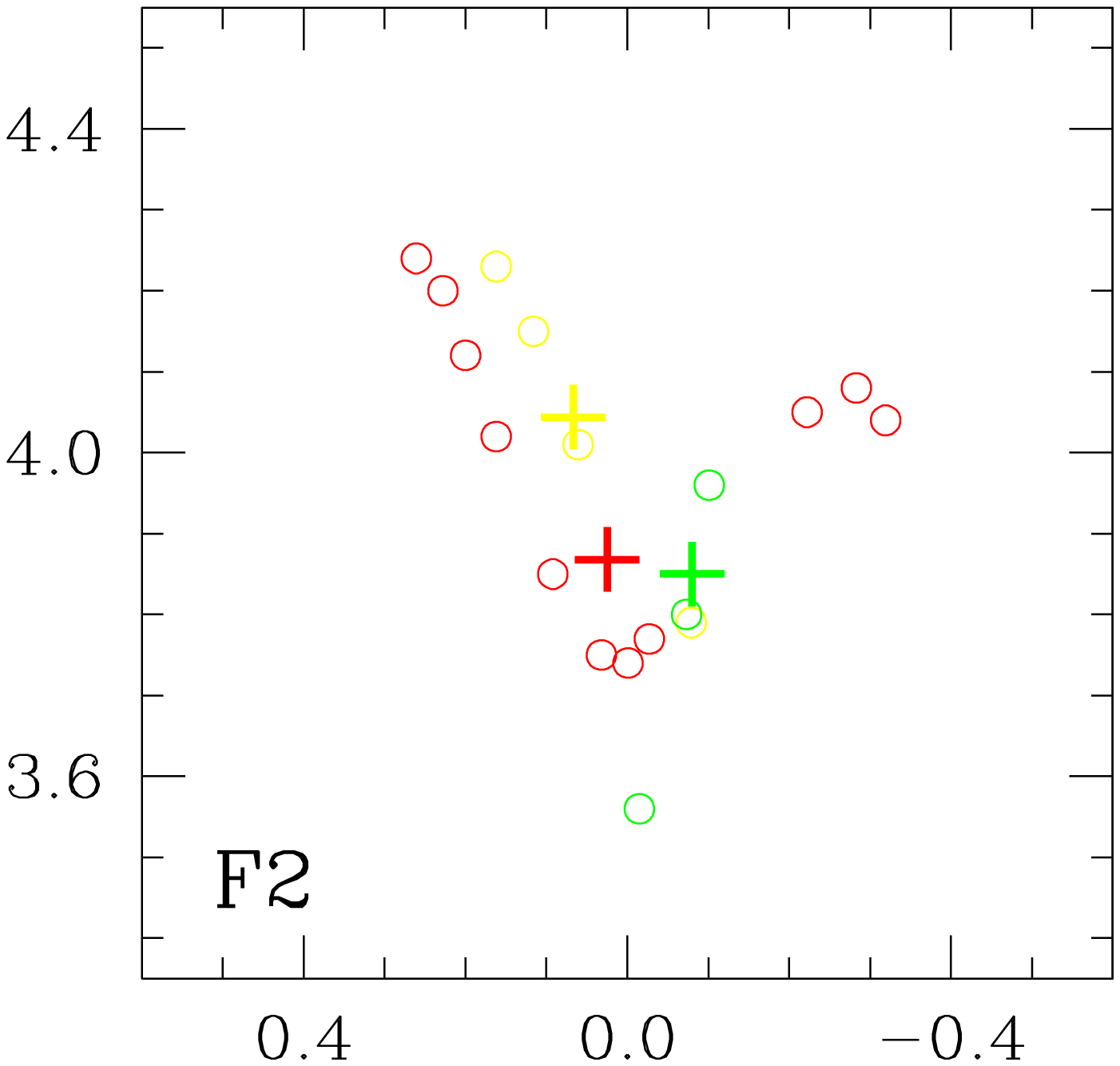}   
    \includegraphics[width=4.2cm,angle=0,clip]{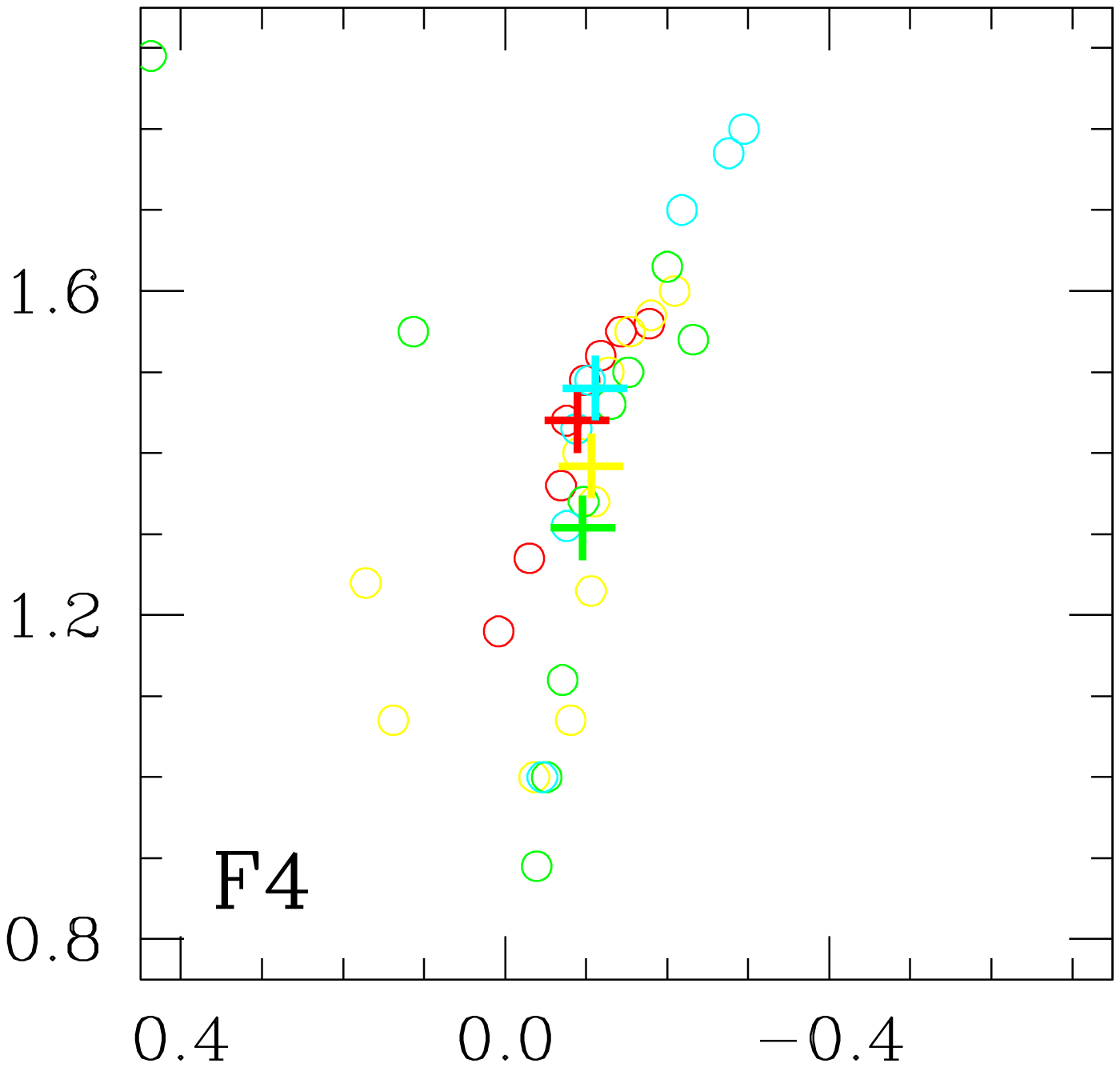}
    \includegraphics[width=4.2cm,angle=0,clip]{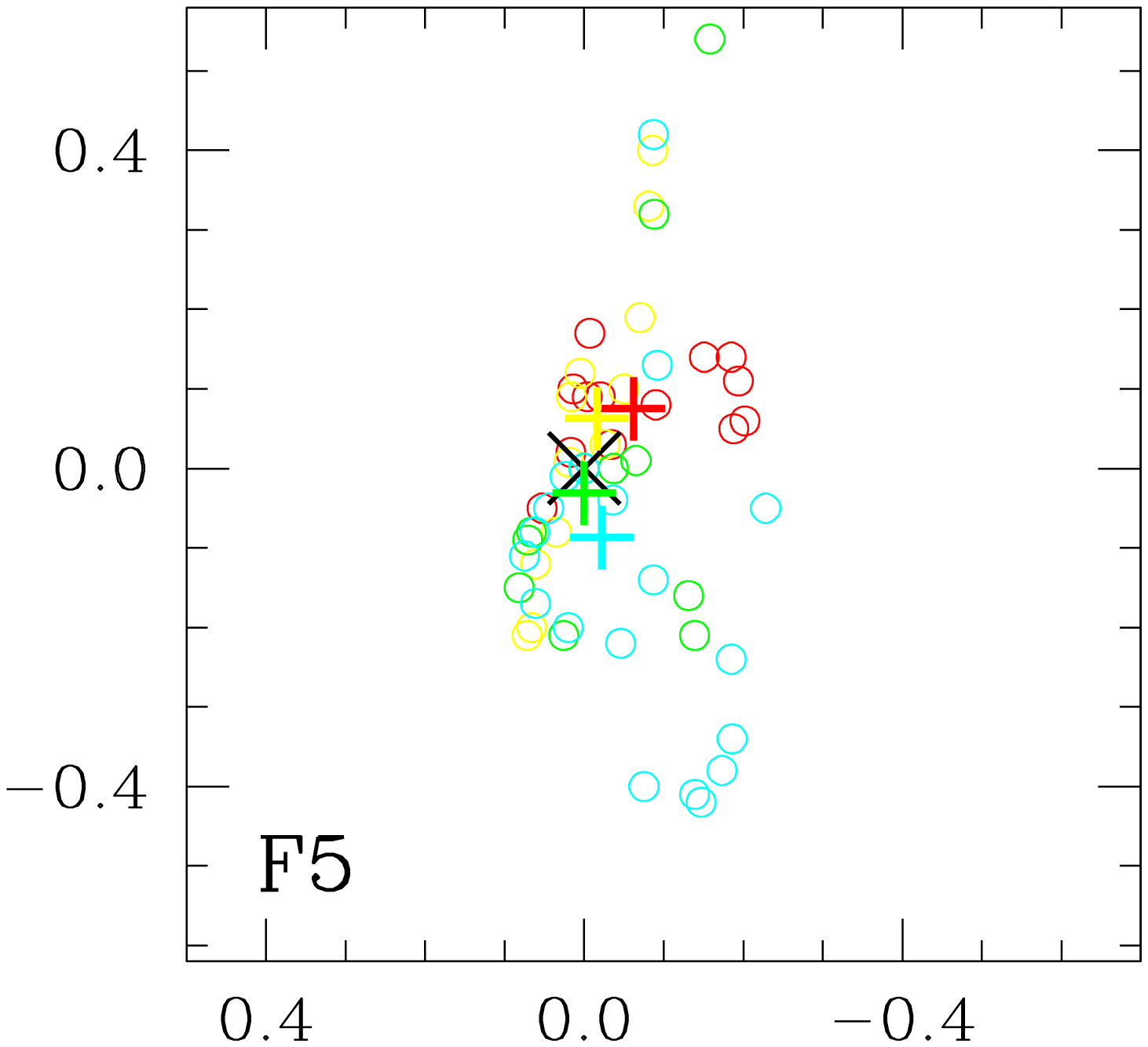} \\
    \includegraphics[width=4.2cm,angle=0,clip]{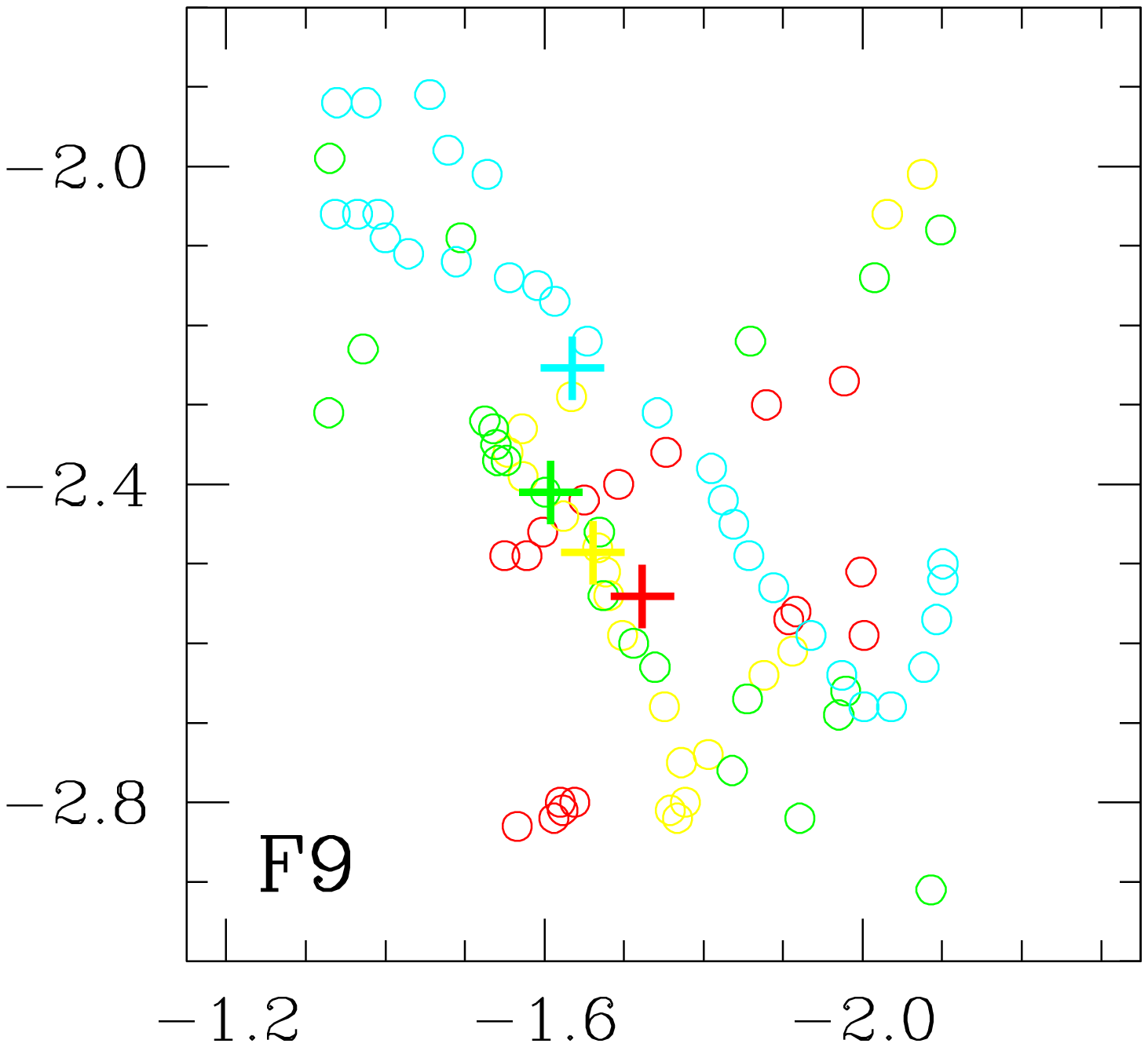}
    \includegraphics[width=4.2cm,angle=0,clip]{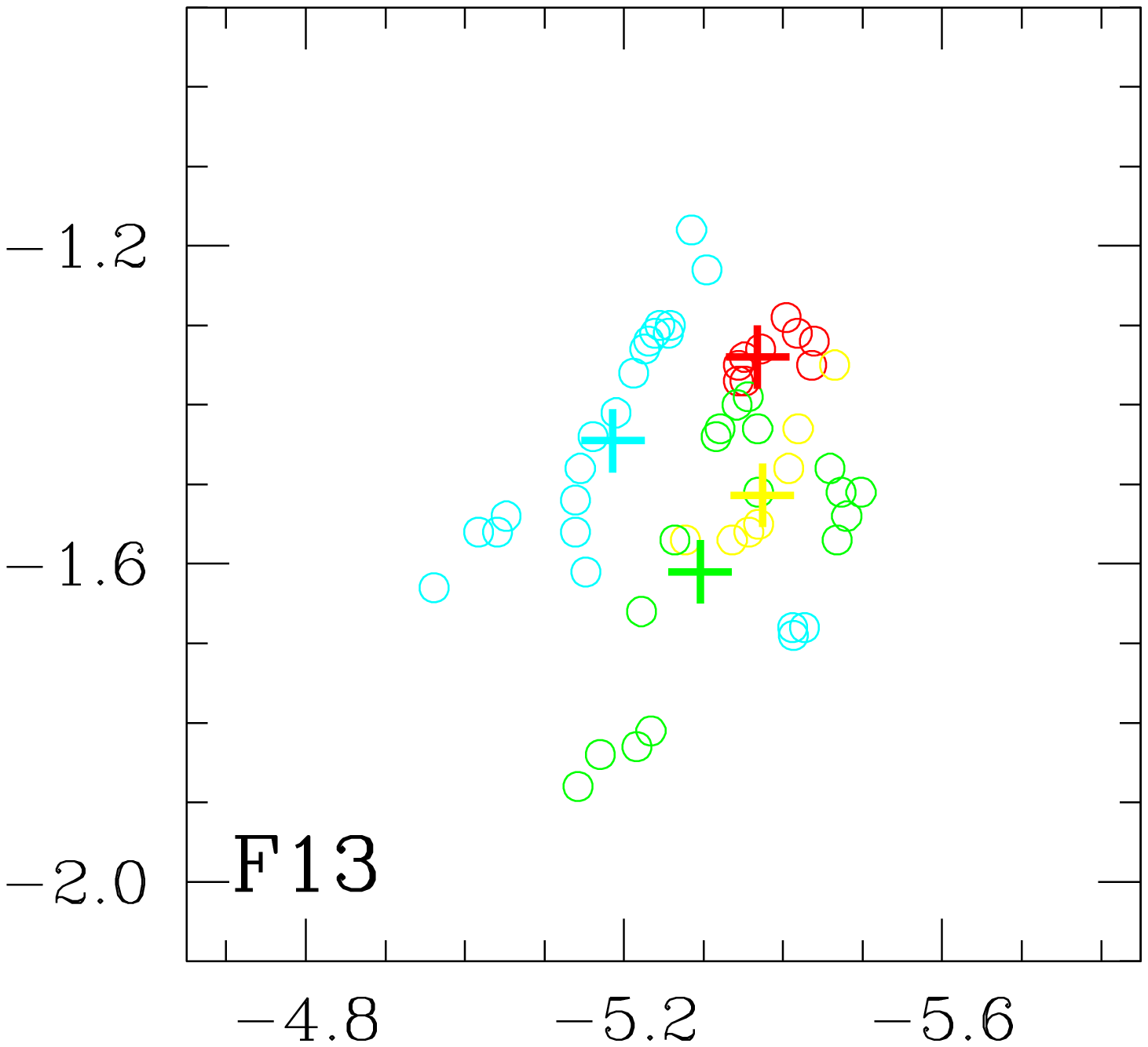} \\
    \includegraphics[width=8.2cm,angle=0,clip]{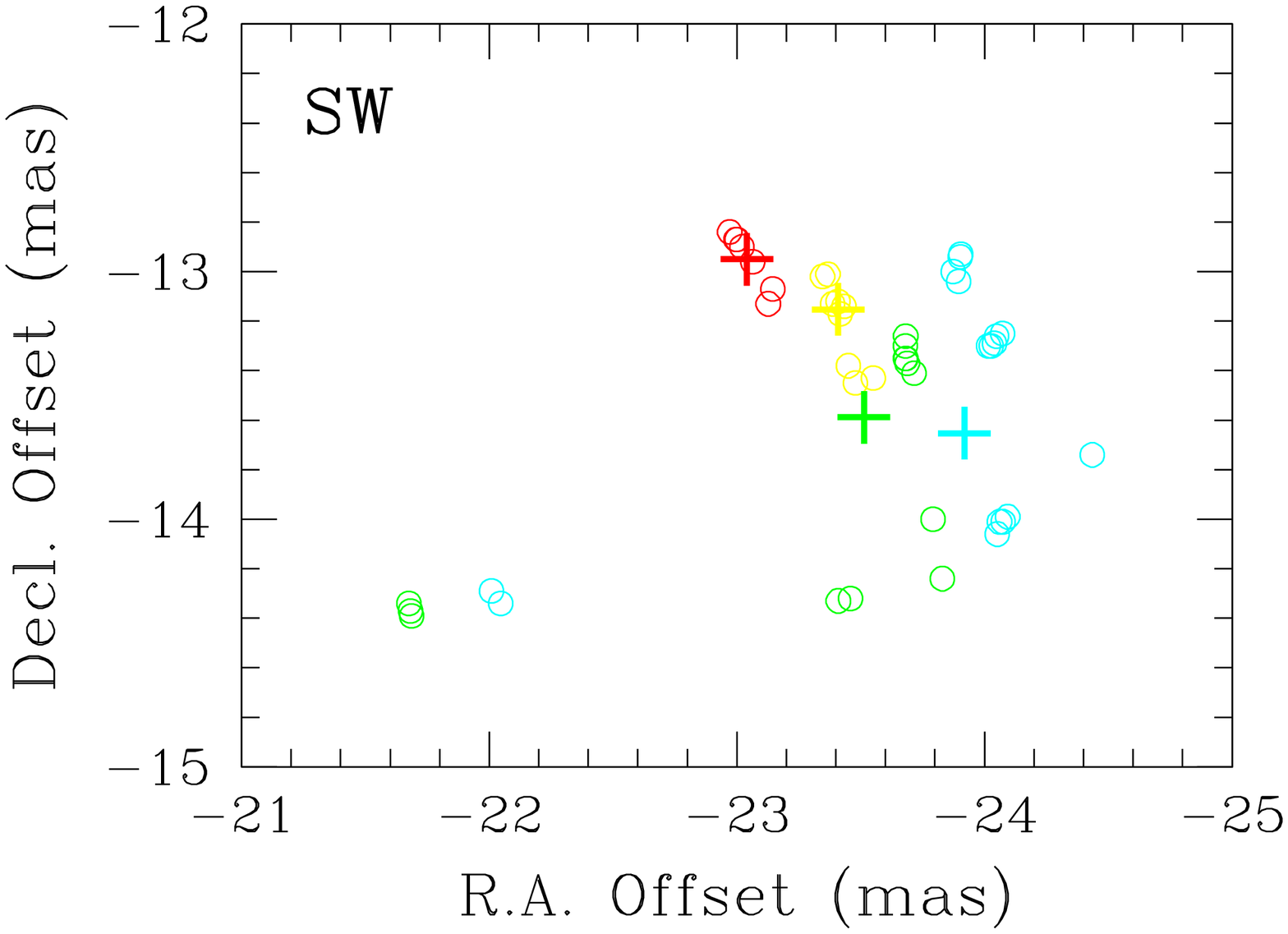}
    \includegraphics[width=5.8cm,angle=0,clip]{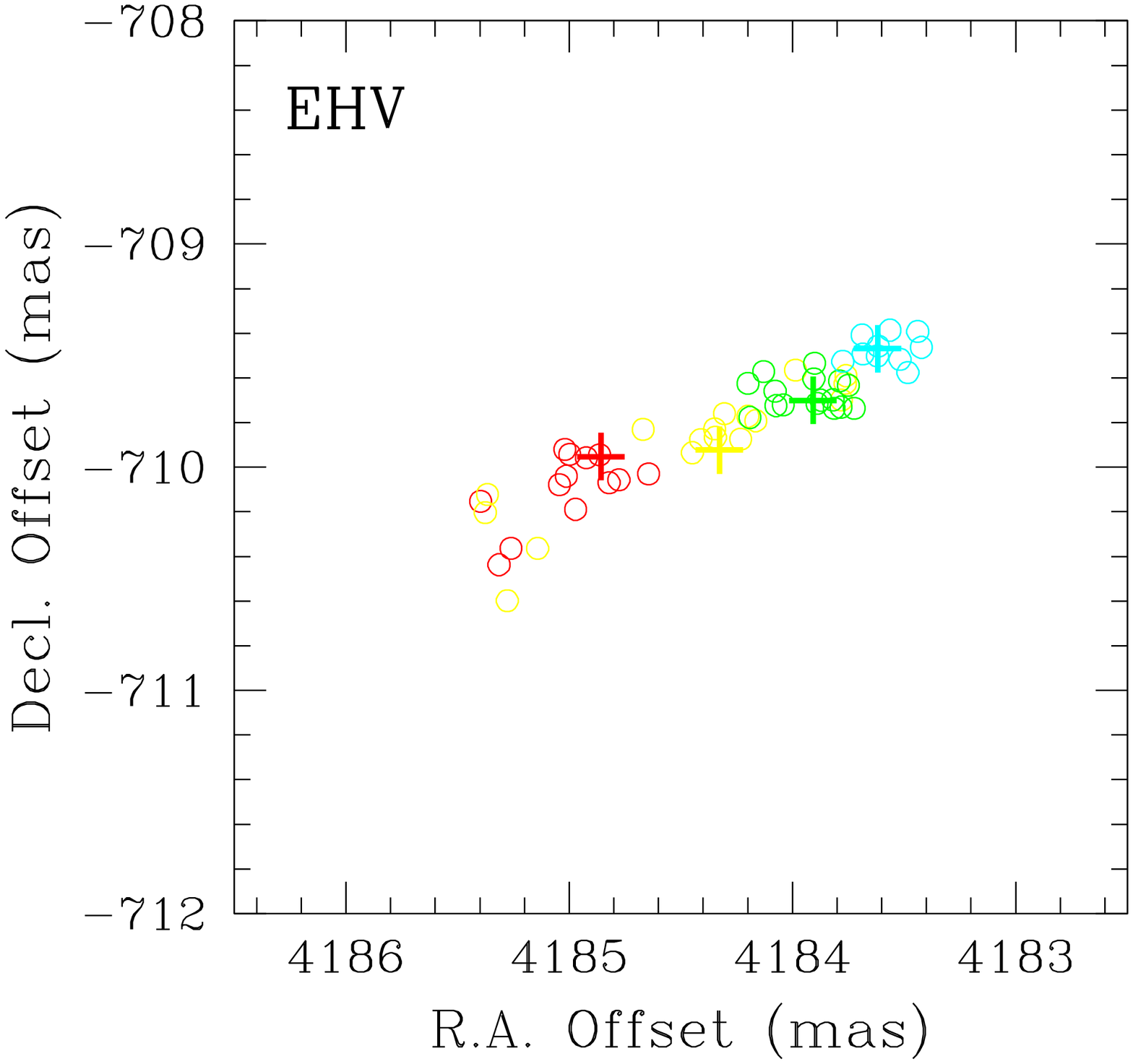}
      \caption{Overlay of the peak positions of the maser spots in 
Features 2, 4, 5, 9, 13, SW and EHV where the maser emission was detected 
continuously more than 3 epochs. 
The colors indicate the observing epochs as in Figure \ref{fig:overlay2}. 
Each color-coded plus mark indicates the intensity-weighted 
mean position of the features at each observing epoch.
The black cross in the Feature 5 indicates the adopted positional
reference spot at $V_{\rm LSR}=$ 1.6 km s$^{-1}$.
              }
         \label{fig:pmfeature}
   \end{figure*}

   \begin{figure*}[ht*]
   \centering  
    \includegraphics[width=3.6cm,angle=-90,clip]{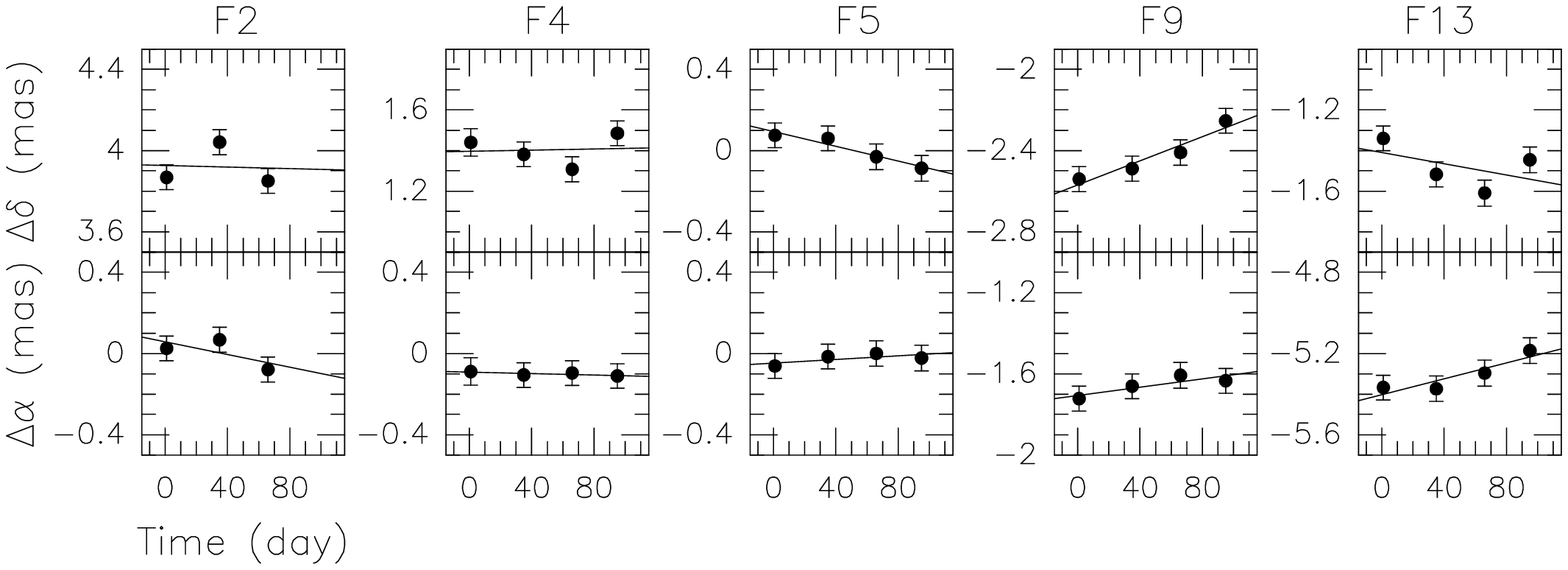} 
    \includegraphics[width=7.2cm,angle=-90,clip]{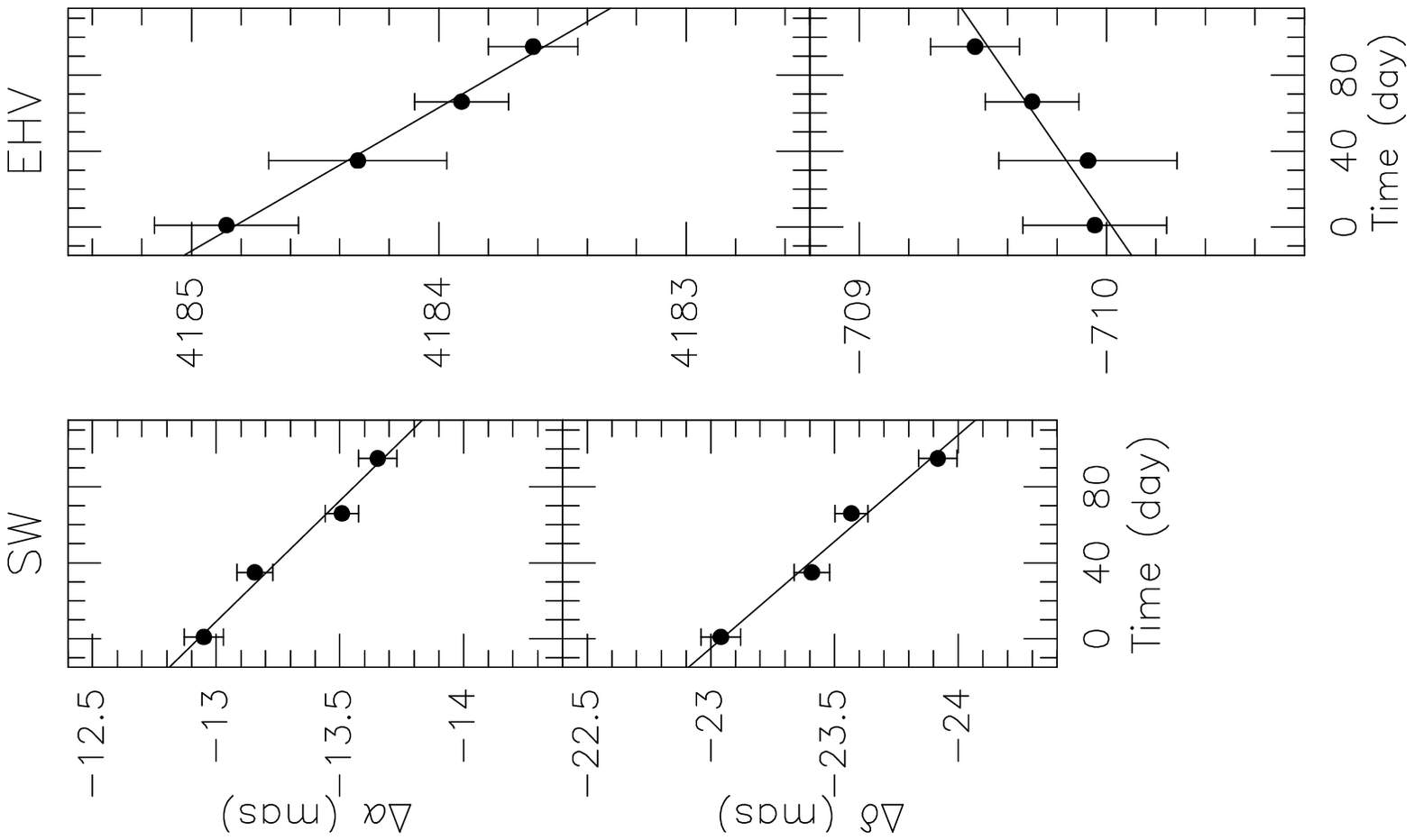}
      \caption{Plots of intensity-weighted mean positions of the 7 features vs. time. 
Error bars include the positional uncertainties in Table \ref{tbl:posfeature}
and the size of the reference spot at 1.6 \kms \ ($\S$\ref{ss:pm}).
The straight lines are the best-fit to the data.
The origin of the time axis is 1999 April 1.
              }
         \label{fig:rdist}
   \end{figure*}

\begin{table*}[ht]
\begin{center}
\newcommand{\lw}[1]{\smash{\lower2.ex\hbox{#1}}}
\caption{Position Shifts and Proper Motions for the 7 Maser Features}
\label{tbl:pm}
\begin{tabular}{lrrrrrrr}
\hline\hline
\lw{Feature} & \multicolumn{1}{c}{$\Delta\alpha^a$} & \multicolumn{1}{c}{$\Delta\delta^b$} & \multicolumn{1}{c}{$|\mu|^c$} & P.A.$^d$ & \multicolumn{1}{c}{$V_{\rm trans}^e$} & \multicolumn{1}{c}{$V_{\rm 3D}^f$} & \multicolumn{1}{c}{$i^g$} \\
   & \multicolumn{1}{c}{(mas)}    & \multicolumn{1}{c}{(mas)}   & \multicolumn{1}{c}{(mas)} & (deg)   & (km s$^{-1}$) & (km s$^{-1}$) & \multicolumn{1}{c}{(deg)} \\ \hline
2$^h$  & $-$0.018$\pm$0.31  & $-$0.15$\pm$0.16 & (0.15$\pm$0.35) & $\cdot\cdot\cdot$ & $\cdot\cdot\cdot$ & $\cdot\cdot\cdot$ & $\cdot\cdot\cdot$ \\
4  &    0.013$\pm$0.13  & $-$0.017$\pm$0.011 & (0.021$\pm$0.14) & $\cdot\cdot\cdot$ & $\cdot\cdot\cdot$ & $\cdot\cdot\cdot$ & $\cdot\cdot\cdot$ \\
5  & $-0.17\pm 0.034$  & 0.042$\pm$0.03 & (0.18$\pm$0.076) & $\cdot\cdot\cdot$ & $\cdot\cdot\cdot$ & $\cdot\cdot\cdot$ & $\cdot\cdot\cdot$ \\
9  & 0.28$\pm$0.058  & 0.098$\pm$0.043 & (0.30$\pm 0.094$) & $\cdot\cdot\cdot$ & $\cdot\cdot\cdot$ & $\cdot\cdot\cdot$ & $\cdot\cdot\cdot$ \\
13 & $-0.13\pm 0.017$  & 0.19$\pm$0.062 & (0.23$\pm$0.06) & $\cdot\cdot\cdot$ & $\cdot\cdot\cdot$ & $\cdot\cdot\cdot$ & $\cdot\cdot\cdot$ \\
SW & $-$0.75$\pm$0.078  & $-$0.85$\pm$0.085 &    1.13$\pm 0.11$ & $-131.4$ & 14.4$\pm$1.4  &   14.4$\pm$1.4 &  4.7$\pm$3.5 \\
EHV & $-$0.50$\pm$0.12  & $+1.26\pm$0.085  & 1.36$\pm 0.16$ &  $-68.2$ & 17.9$\pm$5.2  & $-91.3\pm$29.5 & 79.4$\pm$0.2  \\ \hline
\hline
\end{tabular}
\end{center}
$(a)$ and (b) Total position shifts of the maser features over the observing
period, namely per 95 days, along R.A. and Decl. directions, respectively, 
derived from the least-square fitting in Figure \ref{fig:rdist}.
$(c)$ $|\mu|=\sqrt{(\Delta\alpha)^2+(\Delta\delta)^2}$.
For the maser features where we could not detect well-defined proper motions,
apparent position shifts are given in the parenthesis.
$(d)$ Position Angle of the proper motion vector in the plane of the sky.
$(e)$ Transverse velocity on the plane of the sky converted from $|\mu|$.
$(f)$ 3-Dimensional velocity obtained from 
$V_{\rm 3D}=\sqrt{V_{\rm trans}^2+(V_{\rm mean}-V_{\rm ref})^2}$ where 
$V_{\rm mean}$ is in Table \ref{tbl:spfeature} and
$V_{\rm ref}$ is the LSR-velocity of the reference spot (1.6 km s$^{-1}$).
Negative and positive represent approaching and receding motions, respectively.
$(g)$ Inclination angle of the $V_{\rm 3D}$ vector with respect to the plane of 
the sky, obtained from $i=\tan^{-1}(|V_{\rm mean}-V_{\rm ref}|/V_{\rm trans})$.
$(h)$ All the parameters for Feature 2 are derived from the data taken in the first 3 epochs.
\end{table*}

\subsection{Proper Motions}
\label{ss:pm}

Since VLBI observations generally do not provide absolute positions,
we adopt the brightest spot at $V_{\rm LSR}=1.6$ km s$^{-1}$ in Feature 5
as a reference to examine the cross-epochal positional shifts of the maser features.
The selection of the 1.6 km s$^{-1}$ spot is justified because
this spot does not show any velocity drift over the 4 epochs.
In addition, the spot is enough small to be used as a position reference 
as shown in the following.
The correlated flux (i.e., fringe amplitude) 
of the spot drops by a factor 2 at a projected baseline length 
of $\sim 3.2\times 10^8~\lambda$ which corresponds to a fringe spacing of $\sim 0.06$ mas.
This means that FWHM of the maser spot would be $\sim 0.06$ mas
assuming that the spot has a Gaussian shape brightness distribution.
Note that the reference spot is not a complete point source compared with the
relative position accuracy of the features (Table \ref{tbl:posfeature}).
Therefore, the size of the reference spot will be treated as a
positional uncertainty in our analyses.

In Figure \ref{fig:overlay2}, we present an overlay of the peak positions of 
the maser spots together with boxes indicating the identified features.
The overall distribution of the maser spots does not change over
the 4 epochs:~it shows an arc-shaped structure.
In Figure \ref{fig:pmfeature}, we present magnified overlays of the 
7 features together with the intensity-weighted mean positions for the observing epochs 
(Table \ref{tbl:posfeature}).
As expected from Figure \ref{fig:overlay2}, 
most of the maser features in the MAIN Group do not show any prominent
positional shifts.
On the other hand,
Features SW and EHV displayed distinct positional shifts 
as can be seen in Figure \ref{fig:pmfeature}:
the member spots of SW sequentially appeared from NE to SW
over the 4 epochs and those of EHV appeared from SE to NW.
These facts strongly suggest that the observed position shifts are caused
by real motions of the spatially localized masing gas.
In contrast, Features 2, 4 and 5 show neither well-defined 
position shifts of member spots nor systematic motions of their
intensity-weighted mean positions.

In order to identify proper motions more quantitatively,
we made plots of the intensity-weighted mean positions vs. time (Figure \ref{fig:rdist}).
Table \ref{tbl:pm} summarizes the results of our analysis:~
we showed the derived position displacements for the 7 features.
After fifth column, 
we show results only for the maser features being identified to have proper motions.
We assess that the position shifts seen only in Features SW and EHV represent 
real motions of masing gas because these
features displayed position shifts 
in $|\mu|$ exceeding $3\sigma$ levels of the uncertainties.
In Table \ref{tbl:pm}, we present 
P.A. of the proper motions, 
transverse velocities ($V_{\rm trans}$) in the plane of the sky.
In addition,  we estimated 3-Dimensional (3D) velocities ($V_{\rm 3D}$)
and an inclination angle ($i$) to the plane of the sky.\par

On the other hand, we conclude that the mean position shifts of the 
Features 9 and 13 were not real gas motions, but apparent changes 
because their spot distributions seem {\it random}.
Let us suppose maser appearance reflects the motions of shock fronts 
through a gas clump, as seems likely.  
Then the observations may catch the clump harboring Feature 9 as 
one shock dies out in epoch I (e.g. SE-NW line of maser spots in 
Figure \ref{fig:pmfeature}) and as 
another becomes dominant in epochs II, III, and IV 
(e.g. NE-SW line of maser spots in Figure \ref{fig:pmfeature}).  
In this interpretation, there appears to be little proper
motion between epochs II and III, however
there does appear to be some 
between epochs III and IV.  
Since the status of these aspects of Feature 9 remain unclear,
we do not include them in our analysis of proper motions.  
A dataset with more closely spaced observations, 
extending over a similar time period, could possibly 
clarify the situation.

\subsection{Results of EHV Maser Emission}
\label{ss:ehv}

In $\S$\ref{ss:pm}, we presented that the observed position shift of
the EHV maser emission is the proper motions of masering gas.
It should be noted that the proper motion of the EHV masers is almost directed toward the MAIN 
when we compare Figures  \ref{fig:3mmcont} with \ref{fig:pmfeature}.
Together with its line of sight velocity, we estimate that
3D motion of the masering gas has inclination angle of $79.4^{\circ}\pm 0.4^{\circ}$
with respect to the plane of sky (Table \ref{tbl:pm}), 
suggesting that the gas motion is closely parallel to the line of sight.
Applying the inclination to the apparent angular separation of $\sim 4.2''$
between the MAIN and EHV
(Table \ref{tbl:pm}; corresponding to $\sim 2900$ AU at $d=700$ pc), 
their 3D separation is estimated to be approximately $23\arcsec$
($\simeq 16,000$ AU $\simeq 0.078$ pc).
Even if considering possible uncertainties, 
the real separation is very likely to be in the range between $\sim 2,900$ AU and $\sim 16,000$ AU.\par

We argue that the EHV masers represent a different protostar's activity from 
the Main for the following reasons.
First, the separation is too large to associate the EHV masers with the Main, 
which is likely to be excited by the intermediate-mass protostar ($\S\ref{s:i20intro}$).
In fact,  Furuya et al. (2003) reported that all the twenty \wat\ maser sources detected in their VLA survey 
toward low- and intermediate-mass YSOs were associated with the central protostars within 
$\lesssim 50\sim 200$ AU (see, e.g., Terebey, Vogel, \& Myers 1992).
On the other hand, Hofner \& Churchwell (1996) showed that a median separation between \wat\ masers
and ultra compact \HII\ regions is 0.1 pc for OB stars.
Clearly, these facts support the above conclusion.
Second, if masers are originated in the larger scale EHV CO outflows from the MAIN, 
we could have observed expanding motions.
However, our proper motion measurements clearly showed that their separation has decreased.
On the basis of the large separation and the direction of proper motions,
we thus rule out the hypothesis that EHV masers are associated with the jet emanated from the MAIN.
We suggest that the EHV masers must be associated with another member source in the cluster
(Chen et al. 1997; Wilking et al. 1989).\par

In order to search for the possible exciting source of the EHV
maser emission, we analyzed 4.86 GHz radio continuum emission data from
the VLA Archive Database. We, however, could not detect a compact 
radio continuum emission with a 3$\sigma$ upper limit of 
0.14 mJy beam$^{-1}$ 
($\theta_{\rm maj}\times\theta_{\min}\simeq 5.97''\times 5.15''$).
On the basis of the lack of a bright compact continuum source
and the presence of the EHV maser emission, 
we speculate that the driving source of the
EHV emission could be an extremely young protostar.

   \begin{figure}[ht]
   \centering
    \includegraphics[width=6.7cm,angle=0,clip]{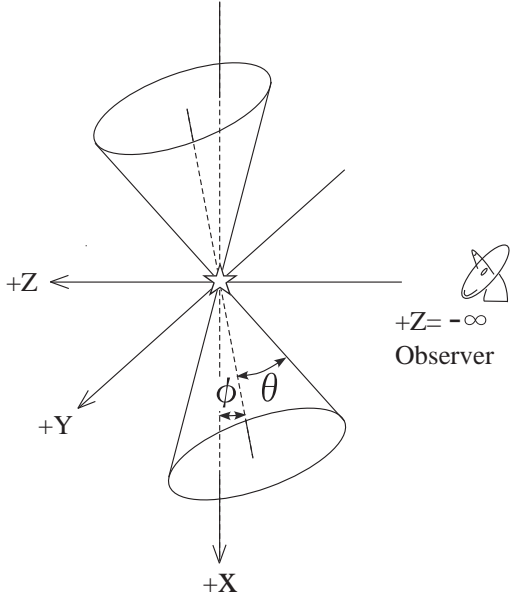}
      \caption{
Sketch of the biconical jet model. 
The $z$ and $x$ axes correspond to the line of sight 
and the jet axis projected on the plane of the sky,
respectively.
An observer lies at $z=-\infty$, and the $x-y$ plane corresponds to the
plane of the sky.
              }
         \label{fig:conemdl}
   \end{figure}

   \begin{figure}[ht]
   \centering
    \includegraphics[width=6.7cm,angle=-90,clip]{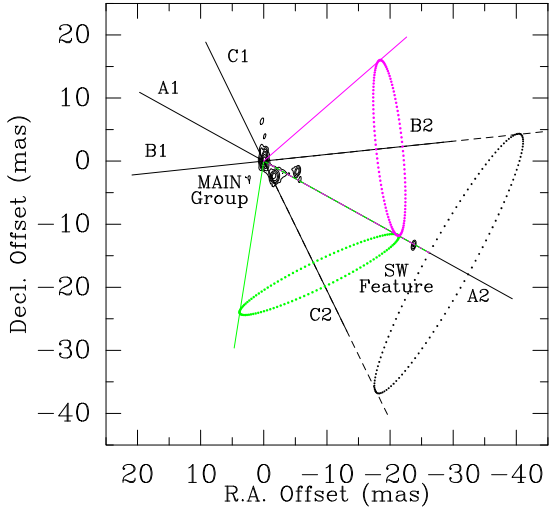}
      \caption{
Guide map showing the 3 jet axes that have been taken as cutting lines
in the position-velocity diagrams (see text; Figure \ref{fig:bestfit}).
To avoid confusion, we present only an outline of each single lobe of the jet.
The position angles of the 3 jet axes in the upper panel are 
$61^{\circ}$ for the axis with the labels of A1 and A2,
$101^{\circ}$ for B1 and B2, and $28^{\circ}$ for C1 and C2.
              }
         \label{fig:pvguide}
   \end{figure}

\section{Discussion}

Evidence from AU-scale VLBI \wat \ maser observations
suggests that the masers in star forming regions are most likely to be excited in the
interaction zone between a jet and ambient cloud material; or in interaction
between a wide angle flow and the surface of a protostellar disk.\par

In this section, we discuss the origin of \wat \ masers associated with the
millimeter continuum source MMS1 in terms of the jet scenario, supported 
by the following evidence.
We do not further discuss the EHV maser emission which we have shown to be
associated with other YSO activity ($\S\ref{ss:ehv}$).
We first note that the spatial and velocity structures of Features 2 to 12
in the MAIN group (Figure \ref{fig:overlay}) convincingly demonstrate that 
they are excited in outflowing gas associated with a protostellar jet.
The maser velocity gradient parallels a line connecting MAIN and SW masers. 
It also parallels the
NE-SW pair of the Intermediate High Velocity (IHV) CO outflow lobes
(P.A.$\simeq -150^{\circ}$; BFT95), although on a smaller scale than the
$10\arcsec$ (corresponding to 0.034 pc) scale of the CO lobes.
Secondly, the velocity sense of the two flow signatures agrees:
the blueshifted masers lie to the NE side and the redshifted masers lie to the SW,
the same as found in the velocity structure of the IHV CO outflow lobe pair.
Third, the relative proper motion between the MAIN and SW shows expansion,
and its direction (P.A.$=-131^{\circ}$; Table \ref{tbl:pm}) is almost parallel 
to the flow line.
These results strongly suggest that the maser jet channels along the direction of the
line connecting MAIN and SW.
It would be difficult to reconcile these motions with interaction of a flow
with a protostellar disk.\par

We assume that the exciting source of the jet is located 
very near the position of the reference 1.6 \kms \ spot for the following reasons.
First, Feature 5, which hosts the 1.6 \kms \ spot, shows a single-peaked spectrum
over all 4 epochs (Figure \ref{fig:sp}), and it did not show 
a velocity drift (Table \ref{tbl:spfeature}).
Second, 
the intensity weighted mean position of Feature 5 
and position of the 1.6 \kms\ spot showed a positional coincidence 
within 0.11 mas over the 4 epochs (see Figure \ref{fig:pmfeature}), which 
is consistent with the result that the reference spot would have a 
FWHM$\sim 0.06$ mas
derived from the fringe amplitude analysis ($\S\ref{ss:pm}$).
Last, the intensity weighted mean positions lie near to a line connecting 
the MAIN and SW masers, suggesting that the 1.6 \kms \ spot is located at 
the expansion center.\par

   \begin{figure}[ht]
   \centering
   \includegraphics[width=8.2cm,angle=0,clip]{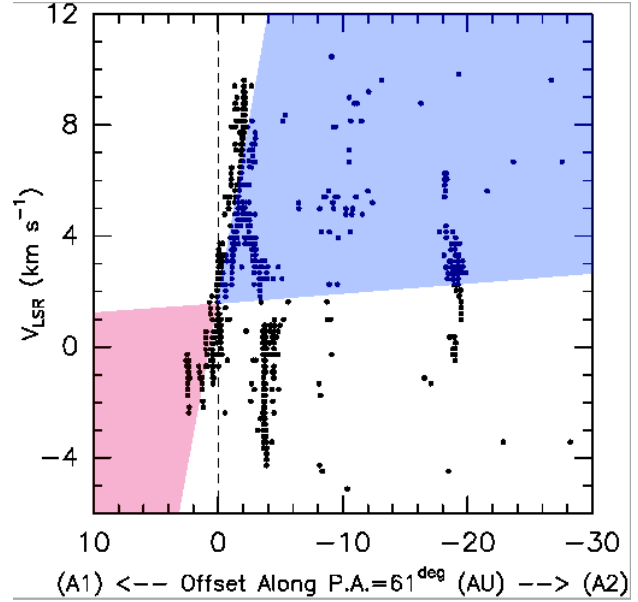}
      \caption{
Comparison of the radial velocities from the biconical jet model 
with the observed LSR-velocities in the position-velocity (PV) diagram
along the A1--A2 axis.
Here we plotted all of the maser spots detected in the 4 epoch observations.
The vertical and horizontal dashed lines, respectively, represent the positions
of the reference 1.6 \kms \ spot.
Expected blue- and redshifted radial velocity regions from the model 
are hatched with blue and red, respectively (see text).  
}
         \label{fig:bestfit}
   \end{figure}

Now we try to shed light on the nature of the masers using
a simple jet model.
We assume that the \wat \ maser emission is excited in the material
at the interface between the protostellar jet and the ambient gas,
namely, at the surface of the cone whose opening angle is $2\theta$.
Such a model has successfully explained the distribution of \wat \ masers 
in the high-mass (proto)star IRAS 20126+4104 \cite{mosca00}.
Figure \ref{fig:conemdl} schematically shows the model: 
the assumed protostar lies at the apex of the cone.
The axis of the cone is inclined by an angle of $\phi$ with
respect to the plane of the sky.
We define a coordinate system whose
$z$ and $x$ axes are, respectively, parallel to the line of sight and 
the projection of the jet axis on the sky. 
To calculate the jet velocities seen by an observer ($v_{\rm z}$) at $z=-\infty$,
we consider that the gas 
moves along straight lines passing through the apex into two opposite 
directions.
Given a power-law velocity profile of $v(r)=v_0(\frac{r}{R})^{\alpha}$, 
the jet velocity along the line of sight can be written as

\begin{equation}
v_{\rm z}(x)=v_0\left\{\frac{x}{R\cos(\phi +\theta )}\right\}^{\alpha}\sin(\phi +\theta )
\end{equation}

Here we adopt $v_0$ of 14.4 \kms\ from the 3D-velocity of the SW maser
(Table \ref{tbl:pm}),
and its distance of 21 AU from the 1.6 \kms \ spot for $R$.
We believe that this assumption is reasonable when we considered other results from
VLBA proper motions studies in low- and intermediate-mass YSOs
(IRAS 05413$-$0104: 64$\pm$22 \kms\ at a distance of 40 AU from the expansion center 
[Claussen et al. 1998], 
S106 FIR:~ 25--40 \kms\ at 25 AU [Furuya et al. 2000],
IRAS 21391$+$5802:~ $\lesssim 42$ \kms\ at $\sim 20$ AU [Patel et al. 2000],
NGC 2071 IRS3:~ 22--42 \kms\ at 260 AU [Seth et al. 2002]).
We thus have the following four free parameters:
P.A. of the projected jet axis to the plane of the sky
(i.e., P.A. of $x$-axis),
$\phi$, $\theta$, and $\alpha$.
By definition, we can give constraints of
$0< \theta <\pi /2$ and $0 <|\phi |< \pi/2$.
To apply such model,  we selected 3 possible axes of 
P.A.$=61^{\circ},~101^{\circ}$, and $28^{\circ}$ which we refer to as, respectively,
A1--A2, B1--B2, and C1--C2 (see Figure \ref{fig:pvguide}).
Note that $61^{\circ}$ is the P.A. of the line connecting
the reference spot and the SW maser, and
that B1--B2 and C1--C2 are parallel to the two pairs of CO outflow lobes 
(see Figure 3 of BFT95).
As for the power-law indices of the jet velocity, 
we considered representative values of $\alpha=0, ~+1,$ and $-1$
which characterize constant velocity, accelerating, and decelerating jets,
respectively.\par

We compared the observed and calculated velocities 
in position-velocity (PV) diagrams:
the black dots in Figure \ref{fig:bestfit} represent the PV distribution of 
the masers along the A1--A2 axis.
Here we took the LSR-velocity of $-1.6$ \kms \ as a mean velocity.
Figure \ref{fig:bestfit} clearly tells us that
the systemic velocity of \Vlsr \ $=6$ \kms \ 
(see the inserted panel in Figure \ref{fig:overall}) which was
derived from the 0.01 pc scale molecular cloud (BFT95)
is not valid for the 10 AU scale maser emitting region.
We started searching for the best-fit parameters from $\alpha$.
We found that only accelerating jet can explain the observed velocity structure
whereas both constant velocity and decelerating jets cannot.
We thus take $\alpha =+1$, leaving us with
two free parameters --- $\phi$ and $\theta$.
We obtained the best-fit parameters of $\phi=-39^{\circ}$, and $\theta=36^{\circ}$.
The blue- and red hatched regions in the 1st and 3rd quadrants of Figure \ref{fig:bestfit}, 
respectively, show the expected LSR-velocity ranges for the blue- and redshifted components.
Although the observed maser emission is not excited all over the expected regions,
a large opening angle such as 
$2\theta\gtrsim 70^{\circ}$ is required to explain the velocity structure 
in the 1st and 3rd quadrants.
This geometry suggests that the line of the sight would match the cone surface 
(that is, $|\theta|+|\phi| \sim \frac{\pi}{2}$).
However, the remaining maser spots in the 4th quadrant, which are Features 
13 and 14
(see Figure \ref{fig:overlay}), 
do not reconcile with the jet model prediction.\par

Since only the accelerating jet explains the velocity structure,
we extended our analysis to the remaining two axes of B1--B2 and C1--C2, 
keeping $\alpha =+1$.
The best-fit parameters for the B1--B2, and C1--C2 cases are 
essentially the same as
those obtained from the A1--A2 in the following three senses:~

\begin{itemize}
\item (i) Requiring a large opening angle 
($2\theta=72^{\circ}$ for B1--B2 and $74^{\circ}$ for C1--C2),
in other words, the P.A. would have an uncertainty of $35^{\circ}$.

\item (ii) The cone is most likely to have a geometry such that the 
surface where masers are excited is nearly parallel to the line of sight 
($|\theta|+|\phi|=78^{\circ}$ for B1--B2, and $80^{\circ}$ for C1--C2),
namely, the SW masers lie at the either of the two edges of the cone.

\item (iii) The LSR-velocities of the spots in Features 13 and 14 cannot be simultaneously 
reproduced with other spots.

\end{itemize}

In addition, we performed further analysis by changing the P.A. through 5$^{\circ}$
increments,
realizing the following points which might jeopardize the above conclusion:~

\begin{itemize}
\item (a) There is a 4 order of magnitude difference in the spatial
scales between our VLBA images and CO (2--1) maps.
\item (b) There is no clear association of CO outflows with known driving source(s)
($\S\ref{ss:mmcont}$; BFT95; Chini et al. 2001).
\item (c) The maser emission does not trace the whole of the outflowing gas.
\end{itemize}

Nonetheless, 
the analysis with 5$^{\circ}$ P. A. steps showed 
that the above results from (i) to (iii) are valid, and
that no other range of P.A. than from $25^{\circ}$ to $110^{\circ}$ can be 
taken.\par

Of the available observations of this region, our high resolution data has
the best opportunity to discern individual sources of outflow. 
We stress that a single jet model cannot
explain the PV structure of the Features 13 and 14 which 
were located at the most western portion of the chain of the MAIN maser features,
but showed a clear velocity gap with respect to the coherent velocity structure from 
the Features 3 to 12 (see Figure \ref{fig:overlay}).
Can they be used to pinpoint additional sources of outflow, 
perhaps associated with the other jet-like CO outflows from the region?
We examined the velocity structure of the discordant spots 
in terms of a ``multiple jet scenario'' 
by applying the above 
``single jet model'' again on ``residual PV diagrams''. 
We subtracted maser spots in the hatched regions in Figure \ref{fig:bestfit}.
In addition, we subtracted all the SW spots, assuming that such spots are
associated with the ``main'' jet.
Hence, the ``residual spots'' consist of all the spots in Features 13 and 14,
and most of the spots in Features 1--3.
``Residual PV diagrams'' were made along position angles incremented by
$30^{\circ}$, including all the CO outflow position angles.
Given the same driving source as the ``main'' jet,
one of the second jet lobes must lie to the NW, 
which might be characterized with a 
P.A.$\sim 150^{\circ}$, 
$2\theta\sim 130^{\circ}$, and
$\phi\sim  60^{\circ}$, but with no symmetrically placed lobe.
We are unable to convincingly associate the residual spots with any particular
flow geometry or additional discrete source.
We speculate that 
the residual spots are being excited in some shocked regions associated with the
``main'' jet.\par

In conclusion, we summarize our jet model analysis
by combining the results from i) to iii) with those from the 
``residual PV'' analysis:~

\begin{enumerate}

\item The majority of the \wat \ masers are most likely 
to be associated with an accelerating biconical jet 
whose projected axis has a P.A. of $60^{\circ}\pm 35^{\circ}$, and
whose opening angle is larger than about $70^{\circ}$.

\item The cone surface enclosing the jet almost matches the line of sight.

\item We could not clearly explain the origin of remaining maser spots,
and speculate that they are being excited in more complex jet-cloud interactions.

\end{enumerate}

We suggest that MMS1 is probably a single star and is the source of the
flow exciting the masers lying near it.
If there has been no precession(s) of the maser jet and CO outflow(s),
our results indicate that the maser jet is most closely
associated with the NE--SW 
pair of the IHV CO lobes because their axes are almost parallel.
The large opening angle of $\sim 70^{\circ}$ at a distance of 21 AU
from the assumed expansion center
is similar to that measured in the intermediate-mass YSO of IRAS 21391$+$5802
by Patel et al. (2000)--these authors reported an opening angle
of $\sim 110^{\circ}$ at the 150 AU point from the star.
Interestingly these results are also consistent with those from 
high-mass (proto)stars 
(Shepherd, Claussen, \& Kurtz 2001, and references therein)
rather than low-mass protostars where highly collimated CO outflows
are generally seen (Richer et al. 2000, and references therein).
We posit that higher luminosity 
YSOs tend to have poorly collimated jets, 
and suggest that a similar jet morphology obtains 
for intermediate-mass stars.
The conclusion that the masers accelerate agrees
with observations of other intermediate-mass YSOs
such as S106 FIR (Furuya et al. 1999)
and IRAS 21391$+$5802 (Patel et al. 2000).
In addition, this conclusion does not disagree with that from
high-mass star forming region W49N where Gwinn, Moran, \& Reid (1992)
reported that expansion velocity of the masering gas has a constant
velocity of $\sim 18$ \kms\ up to 0.1 pc in a distance from the center,
beyond which their velocity increases to more than 200 \kms\ .\par

Given the complexity of the region 
seen in the CO maps (BFT95) and the near-IR image (Chen et al. 1997), 
we believe that sub-arcsecond resolution imaging of the 
CO outflow and continuum emissions with radio interferometers 
will associate jets and outflows with their driving sources.
These interferometric observations will fill the spatial resolution gap
between our AU scale VLBI view of the masers and the 0.01 pc scale
single-dish telescope view of the CO outflows.
As we have mentioned,
\i20\ has displayed \wat\ maser emission at $V_{\rm LSR}=-72$ and $-78$ km s$^{-1}$ 
since 2003 January, which were not seen during our observations ($\S\ref{sss:ehvsearch}$).
Subsequent monitoring observations using the Green Bank 100-m telescope
(A. Wootten, private communication) showed that these emissions have flared
up to $\sim$20 Jy in 2004 October.
Together with sub-arcsecond interferometric observations, further VLBA \wat\ maser
study will help to assess the nature of this multiple 
jet-outflow system, which harbors some of the highest velocity outflowing gas 
in any star forming region known to date.

\section{Summary}

We have performed
a monthly 4-epoch VLBA observations of the \wat \ masers in the intermediate-mass
protostar IRAS 20050+2720 MMS1 together with aperture synthesis observations 
of 3 mm continuum emission with the OVRO array.
The main results of this study are summarized as follows.

\begin{enumerate}

\item From the VLBA images taken with all the 10 antennas, 
we found the two groups of the low-velocity \wat \ maser spots toward the 
bright millimeter continuum emission peak.
One group (the MAIN group) showed intense emission around the cloud velocity,
the other (the SW feature) was located at a projected distance of 18.2 AU south-west of the MAIN.
The OVRO 3-mm images clearly showed that a bulk of millimeter 
continuum emission (MMS1) is associated with the low-velocity masers.

\item Using only the south-western antennas of the VLBA,
we have succeeded in detecting the EHV maser emission blueshifted
by 99 km s$^{-1}$ to the cloud velocity.
The EHV emission is located at 4400 mas in east and
709 mas south of the low-velocity emission.
Considering the large 3D separation between the MAIN and EHV features,
and the proper motion of the EHV toward the MAIN,
we concluded that the EHV emission is not associated with the low-velocity emission.
A distinct millimeter continuum source appears to be associated with the
EHV masers, which 
we speculate is the driving source of the
EHV emission and which could be an extremely young protostar.

\item The overall structure of the masers in the MAIN group did not 
change over the 4 epochs.
On the other hand, 
we found that the projected separation between the MAIN group and the
SW feature increased by 1.13 mas over the 4 epochs, 
which corresponds to a transverse velocity of 14.4 \kms\ .
This increment of the separation indicates proper motion(s)
of spatially localized masing gas in the MAIN group and/or the SW Features.

\item From the analysis of the velocity field of the masers,
we conclude 
that the majority of the \wat \ masers in \i20 \ MMS1 are likely to be 
associated with an accelerating biconical jet whose opening angle is 
approximately $70^{\circ}$ at a distance of 21 AU from the central star.
The presence of such accelerating jet indicates that
the central protostar is driving the powerful jet.
Moreover, the obtained large opening angle of the jet would support 
the hypothesis that poor jet collimation is an inherent property of luminous YSOs.

\end{enumerate}

\begin{acknowledgements}
We are grateful to all of the staff at the VLBA, VLA, and OVRO.
R.S.F. thanks C. M. Walmsley for discussion and encouragement.
R.S.F. was supported by postdoctoral fellowship program at 
INAF, Osservatorio Astrofisico di Arcetri, Italy.
Research at the Owens Valley Radio Observatory is supported 
by the National Science Foundation through NSF grant AST 02-28955.

\end{acknowledgements}

\end{document}